\documentclass[aps,prd,preprint,preprintnumbers,groupedaddress,nofootinbib]{revtex4}
\usepackage{graphicx}
\usepackage{latexsym}
\usepackage{longtable}

\newcommand{\beq}{\begin{eqnarray}}
\newcommand{\eeq}{\end{eqnarray}}

\newcommand{\pbp}{\langle \bar{\psi} \psi \rangle}

\begin{document}

\title {Two flavor QCD and Confinement}

\author{Massimo D'Elia}
\email{delia@ge.infn.it}
\affiliation{Dipartimento di Fisica dell'Universit\`a di Genova and INFN,
Sezione di Genova, Via Dodecaneso 33, I-16146 Genova, Italy}
\author{Adriano Di Giacomo}
\email{digiaco@df.unipi.it}
\affiliation {Dipartimento di Fisica dell'Universit\`a di Pisa and INFN, 
Sezione di Pisa, largo Pontecorvo 3, I-56127 Pisa, Italy}
\author{Claudio Pica}
\email{pica@df.unipi.it}
\affiliation{Dipartimento di Fisica dell'Universit\`a di Pisa and INFN, 
Sezione di Pisa, largo Pontecorvo 3, I-56127 Pisa, Italy}

\begin{abstract}
We argue that the order of the chiral transition for $N_f=2$ is a
sensitive probe of the QCD vacuum, in particular of the mechanism of
color confinement.  A strategy is developed to investigate the order
of the transition by use of finite size scaling analysis.  An in-depth
numerical investigation is performed with staggered fermions on
lattices with $N_t=4$ and
\mbox{$N_s=12,16,20,24,32$} and quark masses $am_q$ ranging from
0.01335 to 0.307036.  The specific heat and a number of susceptibilities
are measured and compared with the expectations of an $O(4)$ second
order and of a first order phase transition. A second order transition
in the $O(4)$ and $O(2)$ universality classes are
excluded. Substantial evidence emerges for a first order transition. A
detailed comparison with previous works is performed.
\end{abstract}
\preprint{IFUP-TH/2005-10, GEF-TH-2005-02}
\maketitle

\section{Introduction}

\begin{figure}[b!]
\includegraphics*[width=0.7\textwidth]{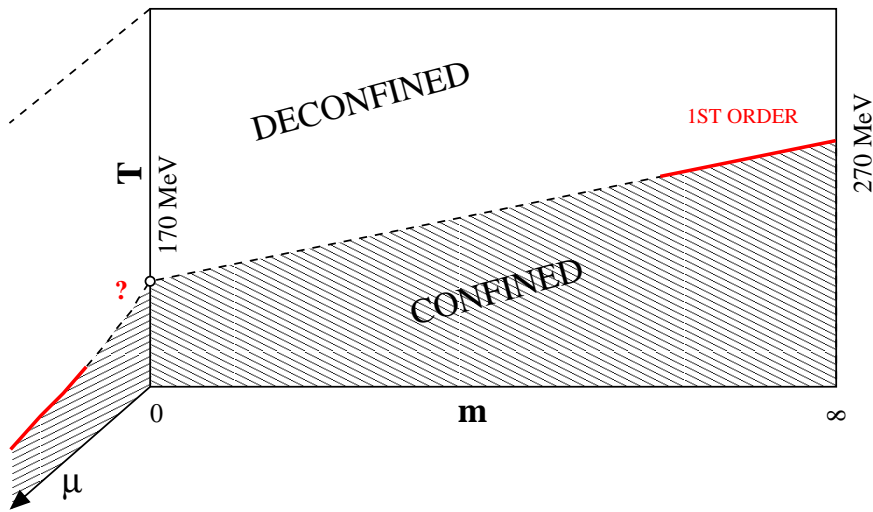}
\caption{Schematic phase diagram of $N_f=2$ QCD.}\label{PHDIA}
\end{figure}

$N_f = 2$ QCD can provide fundamental insight into the mechanism of
confinement.  A schematic view of the phase diagram is shown in
Fig.~\ref{PHDIA}~\cite{Review}.  The quark
masses are assumed to be equal for the sake of simplicity: $m_u = m_d
= m$; $\mu$ is the baryon chemical potential.

Consider the plane $\mu=0$. As $m \to \infty$ quarks decouple and the
system tends to the quenched limit. There the deconfining transition
is well understood: the transition is an order-disorder first order
phase transition, the symmetry involved is $Z_3$ and the Polyakov line
$\langle L \rangle$ is an order parameter. In the presence of quarks
$Z_3$ is explicitely broken and $\langle L \rangle$ is not a good
order parameter. Empirically, however, it works as an order parameter
at quarks masses down to $m \simeq 2.5 - 3$ GeV.

At $m \simeq 0$ a chiral phase transition takes place at $T_c\simeq
170$ MeV, from the low temperature phase where chiral symmetry is
spontaneously broken to a phase in which it is restored: the chiral
condensate $\pbp$ is the corresponding order parameter.  At some
temperature $T_A\geq T_c$ also the $U_A(1)$ symmetry, which is broken
by the anomaly, is expected to be effectively restored.

It is not understood what the chiral transition has to do with the
deconfining transition, but empirically the Polyakov line has a rapid
increase at the transition temperature, indicating deconfinement.

More generally the transition line in Fig.~\ref{PHDIA} is defined by
the maxima of a number of susceptibilities ($C_V$, $\chi_{m}$,
$\dots$) which all coincide within errors, and which indicate a rapid
variation of the corresponding parameters across the line.

A renormalization group analysis plus $\epsilon$-expansion techniques
can be made at $m \simeq 0$, assuming that the relevant degrees of
freedom for the chiral transition are scalar and pseudoscalar
fields~\cite{wilcz1,wilcz2,wilcz3}, or more precisely that the order
parameters are the vacuum expectation values (v.e.v.) of the following fields
\begin{equation}
\tilde\phi: \ \ \phi_{ij} \equiv \langle \bar q_i (1+\gamma_5) q_j\rangle \ \ \ \ (i,j = 1,\ldots,N_f) \, .
\end{equation}
Under chiral and $U_A(1)$ transformations of the group $U_A(1)\otimes
SU(N_f)\otimes SU(N_f)$, $\tilde\phi$ transforms as
\begin{equation}
\tilde\phi \rightarrow e^{i\alpha} U_+ \tilde\phi U_-
\end{equation}
so that by the usual symmetry arguments, and neglecting irrelevant
terms
\begin{equation}
{\cal L}_\phi = \frac{1}{2} Tr \{ \partial_\mu\phi^\dagger\partial^\mu\phi \} - 
\frac{m_\phi^2}{2} Tr \{ \phi^\dagger\phi\} - \frac{\pi^2}{3} g_1 
\left( Tr \{ \phi^\dagger\phi \}\right)^2 - \frac{\pi^2}{3} g_2 Tr \{ (\phi^\dagger\phi)^2\}
+ c\left[ {\rm det}\phi + {\rm det}\phi^\dagger\right] \, .
\label{chiraleff}
\end{equation}
The last term describes the anomaly: indeed it is $SU(N_f)\otimes
SU(N_f)$ invariant, but not $U_A(1)$ invariant.

A second order phase transition corresponds to an infrared (IR) stable
fixed point. For $N_f \geq 3$ the effective action is of the form
Eq.~(\ref{chiraleff}) and no such point exists so that the transition
is first order. For $N_f = 2$, ${\rm det} \phi$ has mass dimension 2
so that other relevant terms emerge, like $({\rm det} \phi + {\rm det}
\phi^\dagger )^2$ and  $Tr \{ \phi^\dagger\phi\} ({\rm det} \phi + {\rm det}
\phi^\dagger )$. If the anomaly term in Eq.~(\ref{chiraleff})
vanishes $(c\simeq 0)$, i.e. if the $\eta '$ mass vanishes at $T_c$,
then there is no IR stable fixed point and the transition is first
order. If instead $c \neq 0$ the symmetry is $SU(2) \otimes SU(2)
\simeq O(4)$ and a fixed point exists which can produce a second order
phase transition.

In the first case the phase transition is first order also at $m \neq
0$ and most likely up to $m = \infty$.

In the second case a phase transition is only present at $m = 0$,
which goes into a continuous crossover as $m \neq 0$: this is true
also in presence of a small chemical potential $\mu \neq 0$, so that a
tricritical point is expected in the $T$-$\mu$ plane (see
Fig.~\ref{PHDIA}) at the border of the crossover with the first order
transition line which takes place in the small $T$, large $\mu$
region~\cite{Stephanov}. Proposals exist to detect the tricritical
point in heavy ion collisions: obviously no such point exists if the
transition at $\mu = 0$ is first order.

The issue is in fact fundamental. If confinement is an absolute
property of the QCD vacuum and the deconfinement transition
corresponds to a change of symmetry (order -- disorder), then a
crossover is excluded and the only allowed possibility is that the
transition is always first order.  The argument also extends to the
case of $2 + 1$ flavors, of which Fig.~\ref{PHDIA} is a
boundary\cite{colombia}. The question deserves a careful study.

A few groups have investigated the problem on the lattice with
staggered~\cite{fuku1,fuku2,colombia,karsch1,karsch2,jlqcd,milc} or
Wilson~\cite{cp-pacs} fermions. The strategy used has either been to
look for signs of discontinuity at the transition, or to study the
dependence on $m$ of the peak of different susceptibilities, or to
study the magnetic equation of state. No clear sign of discontinuity
has been observed, but also no conclusive agreement of scaling with
$O(4)$ critical indexes. In particular the thermal exponent $y_t = 1/\nu$ (see
Sect.~\ref{STRATEGY} for the definition) determined using staggered
fermions differs significantly from that of $O(4)$-$O(2)$ (no direct
determination of the critical exponents exists for Wilson quarks). A
general tendency exists however in the community to consider the
chiral transition second order, and the line of Fig.~\ref{PHDIA} a
crossover.

In the present work we have made a big numerical effort and used large
lattices attempting to clarify the issue. Like most of the other works
we use non improved Kogut--Susskind action, and lattices $4 \times
L_s^3$ with $L_s = 16,20,24,32$. Some scaling violations are expected
and a more careful study with $L_t = 6$ and an improved action is
planned in order to control them.

A preliminary account has been presented at
conferences~\cite{D'Elia:2004ua}. The present paper contains more data
and a full analysis.

The paper is organized as follows: In Section~\ref{STRATEGY} we
explain the strategy used to attack the problem. Section~\ref{NUMERIC}
contains details about the simulations and the numerical results. 
Section~\ref{ANALYSIS} contains the analysis of scaling.
Section~\ref{CONCLUSION} contains a discussion and the conclusions.

\section{Strategy}\label{STRATEGY}

The theoretical tool to investigate the order of a phase transition is
finite size scaling~\cite{FISHER72,BREZIN82}. The extrapolation from
finite size $L_s$ to the thermodynamical limit $L_s = \infty$ is
governed by the critical indexes, which identify the order and the
universality class of the transition.

Approaching the transition, for a higher order or weak first order
transition, the correlation length of the order parameter $\xi$ goes
large compared to the lattice spacing $a$, so that the dependence of
physical quantities on $a/\xi$ can be neglected. More precisely, if
${\cal L}/kT$ is the effective action (density of free energy)
\beq
\frac{\cal L}{kT} \simeq L_s^{-d} \phi \left(\frac{a}{\xi},  \frac{L_s}{\xi}, am_q L_s^{y_h} \right)
\label{scal1}
\eeq
the dependence on $a/\xi$ disappears as $T_c$ is
approached, since $\xi$ diverges as
\beq
\xi \simeq_{\tau \to \infty} \tau^{-\nu}
\eeq
where $\tau \equiv 1 - \frac{T}{T_c}$. The variable $L_s/\xi$ can be traded
with $\tau L_s^{1/\nu}$ and the scaling law follows
\beq
\frac{\cal L}{kT} \simeq L_s^{-d} \phi \left(\tau L_s^{1/\nu}, am_q L_s^{y_h} \right) \, .
\label{scal2}
\eeq

The problem has two scales, $\xi$ and $1/m_q$. The effective action
depends on the order parameter, as dictated by the symmetry, and as
$\tau \to 0$ irrelevant terms can be neglected.  The thermodynamics is
described by correlators of the order parameter, which contain
information on the discontinuities of the thermodynamical quantities.
The most fundamental quantity is the specific heat, which is always
guaranteed to exhibit the correct critical behaviour, independently
of the identification of the correct order parameter.

For the specific heat the scaling law is
\beq
C_V - C_0 \simeq  L_s^{\alpha/\nu} \phi_c \left(\tau L_s^{1/\nu}, am_q L_s^{y_h} \right) \, ;
\label{scalcal}
\eeq
$C_0$ stems from an additive renormalization~\cite{BREZIN82}. 

For the susceptibility $\chi$ of the order parameter $O(x)$
\beq
\chi = \int d^3 x \left[ \langle O(x) O(0) \rangle - \langle O \rangle^2 \right]
\eeq
the scaling law is
\beq
\chi \simeq L_s^{\gamma/\nu} \phi_\chi \left(\tau L_s^{1/\nu}, am_q L_s^{y_h} \right) \, .
\label{scalord}
\eeq
We shall discuss the question if a subtraction is needed for $\chi$
as for the specific heat in Sect.\ref{mag}.

Analogous scaling laws can be derived for mixed susceptibilities.

The values of the indexes characterize the transition: the values
relevant to the analysis which follows are listed in
Table~\ref{CRITEXP}. $O(4)$ is the symmetry expected if the chiral
transition is second order, but it can break down to $O(2)$ by the
lattice discretization for Kogut--Susskind fermions~\cite{karsch1} at
non zero lattice spacing.
\begin{table}[bt!]
\begin{tabular}{|c|c|c|c|c|c|c|c|}
\hline & $y_t$ & $y_h$ & $\nu$ & $\alpha$ & $\gamma$ & $\beta$ & $\delta$\\
\hline $O(4)$ & 1.336(25) & 2.487(3) & 0.748(14) & -0.24(6) & 1.479(94) & 0.3837(69) & 4.852(24)\\
\hline $O(2)$ & 1.496(20) & 2.485(3) & 0.668(9) & -0.005(7) & 1.317(38) & 0.3442(20) & 4.826(12)\\
\hline $MF$ & $3/2$ & $9/4$ & $2/3$ & 0 & 1 & 1/2 & 3\\
\hline $1^{st} Order$ & 3 & 3 & $1/3$ & 1 & 1 & 0 & $\infty$\\
\hline
\end{tabular}
\caption{Critical exponents.}\label{CRITEXP}
\end{table}

The scaling law in Eq.~(\ref{scalcal}) for the specific heat is valid
independent of the knowledge of the order parameter. The scaling law
in Eq.~(\ref{scalord}) instead is correct only if the choice of the
order parameter is the right one. In principle the matching between
(\ref{scalcal}) and (\ref{scalord}) can be used to legitimate any
guess on the symmetry and on the order parameter.

The scaling laws~(\ref{scalcal}) and~(\ref{scalord}) are difficult to
test because they depend on two variables. A possible strategy is to
keep one of them fixed and to study the scaling with respect to the
other. As one can see in Table~\ref{CRITEXP}, the index $y_h$ is the
same within errors for $O(4)$ and $O(2)$ symmetry. In order to reduce
the problem to one scale, we have made a number of simulations at
different values of $L_s$ and $am_q$ keeping $am_q L_s^{y_h}$ fixed
and assuming $y_h=2.49$ which corresponds to $O(4)$ or $O(2)$. In this
way as $L_s$ is increased, $am_q \to 0$, so that the infinite volume
limit corresponds to the chiral transition at $am_q = 0$.

From Eq.s (\ref{scalcal}) and (\ref{scalord}) it follows that the
maxima at constant $am_q L_s^{y_h}$ scale as
\beq
(C_V - C_0)_{\rm max} &\propto& L_s^{\alpha/\nu} \nonumber \\
\chi_{\rm max} &\propto& L_s^{\gamma/\nu} \, .\label{scalmax}
\eeq
as $L_s \rightarrow \infty$ and their positions scale as
\beq
\tau L_s^{1/\nu} = {\rm const}
\eeq
If $O(4)$ or $O(2)$ is the correct symmetry, the values of
$\alpha/\nu$ and $\gamma/\nu$ should be consistent with the
corresponding values listed in Table~\ref{CRITEXP}.

Notice that Eq.s (\ref{scalcal}) and (\ref{scalord}) involve the long
range part of the correlations, i.e. they are related to the infrared
regime, and are not expected to be significantly affected by scaling
violations $O(a/\xi)$.

If the answer to this test is positive the chiral transition is second
order at $am_q = 0$ and a crossover at $am_q \neq 0$. If instead the 
answer is negative and the
assumption of Ref.\cite{wilcz1,wilcz2,wilcz3} about the relevant
degrees of freedom is correct, the transition is first order at $am_q =
0$, and also at $am_q \neq 0$.

An alternative strategy can be as follows. At fixed $am_q$, $\beta$ the
values of the susceptibility should converge at large $L_s$ if ${\cal
L}$ is analytic. One can change variable by replacing $\tau
L_s^{1/\nu}$ with the ratio
\begin{equation}
\frac{\tau L_s^{1/\nu}}{(am_q L_s^{y_h})^{1/(\nu y_h)}} = \tau (am_q)^{-1/(\nu y_h)}\, .
\end{equation} 
The scaling laws Eq.s (\ref{scalcal}) and (\ref{scalord}) become then 
\begin{eqnarray}
C_V - C_0 \simeq  L_s^{\alpha/\nu} \tilde\phi_c \left(\tau (am_q)^{-1/(\nu y_h)},am_q L_s^{y_h} \right)\label{scalcal2A} \\
\chi \simeq L_s^{\gamma/\nu} \tilde\phi_\chi \left(\tau (am_q)^{-1/(\nu y_h)}, am_q L_s^{y_h} \right) \, .
\label{scalord2A}
\end{eqnarray}
At large $L_s$ the dependence on $am_q L_s^{y_h}$ must cancel the
dependence on $L_s$ in front of the scaling functions in
Eq.s (\ref{scalcal}) and (\ref{scalord}). It follows that
\begin{eqnarray}
C_V - C_0 \simeq  (am_q)^{-\alpha/(\nu y_h)} f_c \left(\tau (am_q)^{-1/(\nu y_h)}\right) \label{scalcal1} \\
\chi \simeq  (am_q)^{-\gamma/(\nu y_h)} f_\chi \left(\tau (am_q)^{-1/(\nu y_h)}\right) \, .\label{scalord1}
\end{eqnarray}
The peaks of $(C_V - C_0)$ and of $\chi$ should then scale as
\beq
(C_V - C_0)_{\rm max} &\propto&  (am_q)^{-\alpha/(\nu y_h)} \nonumber \\
\chi_{\rm max} &\propto& (am_q)^{-\gamma/(\nu y_h)}
\label{scalmax1}
\eeq
as $am_q\rightarrow 0$.
As for the position of the maxima, it scales according to
\beq
\tau (am_q)^{-1/(\nu y_h)} = {\rm const} \label{pcscale} \, .
\eeq
An alternative possibility is to keep the scaling in the form
of Eq.s~(\ref{scalcal}) and (\ref{scalord}), and require that the
volume
dependence disappears at $\tau L_s^{1/\nu}$ fixed. This kind of
scaling could work better if the correlation length is comparable
to $L_s$, while $a L_s m_\pi \gg 1$. This implies the scaling laws:
\begin{eqnarray}
C_V - C_0 \simeq  (am_q)^{-\alpha/(\nu y_h)} f_c \left(\tau  L_s^{1/\nu} \right)
\label{scalcal2} \\
\chi \simeq  (am_q)^{-\gamma/(\nu y_h)} f_\chi \left(\tau  L_s^{1/\nu} \right) \, .
\label{scalord2}
\end{eqnarray}
Eq.s~(\ref{scalmax1}) for the maxima stay unchanged, but the positions
of the maxima scale now as 
\beq
\tau L_s^{1/\nu} = {\rm const} \label{pcscale1} \, 
\eeq
and the width of the peaks are volume dependent.

All that is expected to be true at sufficiently large values of 
$a L_s \cdot m_\pi$ and at sufficiently small values of $am_q$, such
that we are not too far from the critical point.

$\tau \equiv 1 - T/T_c$ is usually taken in the literature
\cite{karsch1,karsch2} as proportional to $\beta_0 - \beta$, where
$\beta_0$ is the value of $\beta = 2 N_c / g^2$ at the chiral $(am_q =
0)$ transition, and all the analyses of the scaling law are based on
that choice.

In fact, since 
\beq
T = \frac{1}{L_t a(\beta,am_q)} \, ,
\eeq
the correct definition is 
\beq
\tau \equiv 1 - \frac{T}{T_c} = 1 - \frac{a(\beta_0,0)}{a(\beta,am_q)} \, . \label{taudef}
\eeq
and the dependence on $am_q$ is non trivial (see e.g.~\cite{massapi}).
$a(\beta,am_q)$ is expected to be an analytic function in a neighborhood
of the critical point and therefore for sufficiently small $\beta_c -
\beta$ and $m$
\beq
a(\beta,am_q) \simeq a (\beta_0,0) + \frac{\partial a }{\partial \beta} (\beta_0,0) (\beta - \beta_0) 
+ \frac{\partial a}{\partial (am_q)} (\beta_0,0) am_q \, . 
\label{spacingexp}
\eeq
If needed higher orders in $am_q$ and $(\beta - \beta_0)$ can be
included. It then follows that at sufficiently small values of $am_q$
\beq
\tau = C (\beta_0 - \beta + k_m am_q) \label{taueq}
\eeq
with 
\beq
C &\equiv& \frac{\partial \ln a }{\partial \beta} (\beta_0,0) \nonumber \\
k_m &\equiv& \frac{1}{C}\frac{\partial \ln a }{\partial am_q} (\beta_0,0) \, .
\eeq
In the quenched case this reduces to $\tau \propto \beta_0 - \beta$ as
usual: in the presence of dynamical quarks $k_m \neq 0$~\cite{massapi}.

The scaling law for the position of the peaks becomes then
\beq
\beta_0 - \beta_c + k_m am_q = {\rm const} \cdot (am_q)^{1/(\nu y_h)} \, .
\label{pcbscal}
\eeq
Eq.s (\ref{scalcal1}) and (\ref{scalord1}) should be valid if
$L_s\gg\xi/a$, $L_s\gg 1/(am_\pi)$ (see Table \ref{runpar}). If the
alternative possibility is considered, i.e. requesting that the free
energy stays finite when $L_s\rightarrow \infty$ at fixed $\tau
L_s^{1/\nu}$, the position scales instead as
\beq
\beta_0 - \beta_c + k_m am_q = {\rm const} \cdot L_s^{-1/\nu} \, .
\label{pcbscal1}
\eeq
Analogous formulae can be written including quadratic terms of the 
expansion Eq.~(\ref{spacingexp}) (see Section~\ref{ANALYSIS} below). 

An alternative technique is to investigate the order of the transition
by looking for discontinuities: if the transition is first order at $m
= 0$, it is expected to be so also at $m \neq 0$. If the transition is
weak first order, at small volumes compared to some critical volume it
will behave as if the free energy were regular, so that Eq.s
(\ref{scalcal1}), (\ref{scalord1}) and (\ref{scalmax1}), or
(\ref{scalcal2}) and (\ref{scalord2}), are expected to be valid, with
the critical indexes appropriate to first order. At larger volumes,
however, the peak of the specific heat as well as the peaks of the
other susceptibilities should increase proportionally to the volume,
as a consequence of the discontinuity in the first derivatives of the
free energy. At the same time a bistability should appear in the time
histories~\cite{colombia}. Such an analysis has been in particular
performed in Ref.\cite{jlqcd,milc}: some sign of growth with the
volume has been observed, but no significant bistability; we will
comment on this result below. Of course if a discontinuity is observed
one can conclude that the transition is first order. If not one cannot
exclude that it could be observed at larger volumes.

Finally one can investigate the so-called magnetic equation of
state~\cite{milc}, i.e. the scaling behaviour of the chiral order
parameter itself, $\pbp$, versus the reduced temperature.  The scaling
law is
\beq
\pbp \simeq m^{1/\delta} f(\tau m^{-1/(\nu y_h)})\label{eqstate}
\eeq
and again it can provide information on the critical indexes.

\section{Numerical simulations}\label{NUMERIC}

\subsection{Algorithm}
Monte Carlo simulations were performed using the standard staggered
action. The \textit{Hybrid R} algorithm~\cite{HybridR} was used for
the configuration updating.  Since it is a non-exact algorithm, its
systematic errors must be kept under control.  The finite integration
step used in the molecular dynamics evolution introduces a systematic
error on the mean values of observables proportional to the
integration step squared $\Delta\tau^2$. Great care was taken to
ensure that this systematic shift were much smaller than the
statistical error in each Monte Carlo run.  Typical values of the
integration step $\Delta\tau$ vary with the mass of the quarks in
units of the lattice spacing as $\Delta\tau=a m_q / 4$. When the use
of that value for the integration step was too proibitive, i.e. at the
smallest quarks masses used in this work, the integration step was in
any case taken below $a m_q / 2$.  The stopping condition used for the
conjugate gradient inversion was fixed requiring that the residue were
smaller than $10^{-8}$. The length of the molecular dynamics
trajectories was fixed to $1$ for all of our simulations.

\subsection{Run parameters}

All of our numerical simulations were performed using a lattice
temporal extent of $L_t=4$. To begin with we run two sets of Monte
Carlo simulations fixing for each the value of $am_q L_s^{y_h}$ as
explained in Section~\ref{STRATEGY}. The two sets, called in the
following Run1 and Run2, have $am_q L_s^{y_h}=74.7$
and $am_q L_s^{y_h}=149.4$ respectively. The spatial lattice sizes
$L_s$ used for each of the two sets are $L_s=12, 16, 20, 32$.

Additional simulations at $L_s=24$ and $am_q=0.04444$ and at $L_s=16$
and $am_q=0.01335$ were added.  The second one was chosen by purpose
at the same mass of the $L_s=32$ of Run1.

A summary of the bare quark masses and $L_s$ used is reported in
Table~\ref{runpar}.  The total number of MC trajectories collected is
also reported together with the quantity $aL_s \cdot m_\pi$ at the
pseudocritical value of the coupling $\beta_c$ ($am_\pi$ was taken from
the parametrization given by the MILC collaboration in
Ref.\cite{massapi}).  Since for all our runs the spatial extent is
much larger than the pion correlation length, no large infrared
cut-off effects are expected, except possibly for the run at 
$L_s = 16$ and $am_q = 0.01335$ (see Table~\ref{runpar}).

\begin{table}[b!]
\caption{Run parameters for the numerical simulations.}\label{runpar}
\begin{tabular}{|c||c|c|c|c||c|c|c|c||c|c|}
\hline
& \multicolumn{4}{|c||}{Run1} & \multicolumn{4}{|c||}{Run2} & \multicolumn{2}{|c|}{Other} \\
\hline $L_s$ & 12 & 16 & 20 & 32 & 12 & 16 & 20 & 32 & 16 & 24 \\
\hline $am_q$ & 0.153518 & 0.075 & 0.04303 & 0.01335 & 0.307036 & 0.15 & 0.08606 & 0.0267 & 0.01335 & 0.04444 \\
\hline \# Traj. & 22500 & 87700 & 14520 & 14500 & 25000 & 131390 & 16100 & 15100 & 10000 & 10000 \\
\hline $aL_s \cdot m_\pi$ & 11.9 & 11.0 & 10.0 & 8.9 & 11.3 & 15.8 & 14.8 & 12.4 & 4.5 & 12.2 \\
\hline
\end{tabular}
\end{table}

For each value of $am_q$ and $L_s$ and for each run, MC simulations
were performed at different $\beta$ values in order to inspect and to
have under control the whole interesting critical region. See 
Appendix~\ref{appA} for the whole listing of our run parameters.

\subsection{Data Reweighting}

The collected raw data were analyzed using standard statistical
procedures (see e.g. \cite{newmanbarkema}). For the history of each
observable, thermalizations were taken self consistently to be five
times the integrated autocorrelation time estimated from thermalized
trajectories.

The data collected were analyzed using the multi-histogram reweighting
technique combining data taken at different $\beta$'s together. This
method allows to extract mean values of observables and their
susceptibilities at intermediate $\beta$ values over the whole range
explored with numerical simulations. Using the reweighted data, it is
possible to locate accurately the position at which the
susceptibilities attain their maximum and their value at the maximum.
Sometimes in previous studies a single point with high statistics at
about the critical point was used. Using data from simulations done at
several $\beta$ values covering the whole critical region, eliminates
the risk of a wrong extrapolation from a single $\beta$ too distant
from the critical point\footnote{Remember that for single histogram
reweighting the statistics needed in order to extrapolate measured
quantities at a value of $\beta=\beta_0+\Delta\beta$ different from
that used in the actual simulation grow exponentially with
$\Delta\beta$.}. Moreover the method allows a better sampling of the
probability distribution, due to the fact that different simulations
are combined together, thus increasing the precision and confidence of
the measurement.

The errors of observable quantities were estimated using the bootstrap
method. In practice, this means repeating the whole multi-histogram
reweighting procedure a number of times starting from random data
samples distributed as the measured empirical distributions.

\subsection{Observables}

For each generated configuration of our MC simulations we measured the
average spatial and temporal plaquettes ($P_\sigma$,
$P_\tau$), the chiral condensate ($\bar\psi\psi$), the energy density
($\bar\psi D_0\psi$) and the following lattice susceptibilities (the
notation is the same as in~\cite{jlqcd}):
\begin{eqnarray}
\chi_m^{disc} &=& \left(\frac{N_f}{4}\right)^2 \frac{1}{V}\left[\langle({\rm Tr} D^{-1})^2\rangle-\langle {\rm Tr} D^{-1}\rangle^2\right]\label{LATSUSC1}\\
\chi_m^{conn} &=& -\frac{N_f}{4V} \sum_{x,y} \left<D_{x,y}^{-1}D_{y,x}^{-1}\right>\\
\chi_{e,ij} &=& V [\left<P_iP_j\right>-\left<P_i\right>\left<P_j\right>], \quad i,j = \sigma , \tau\label{LATSUSC3}\\
\chi_{e,f} &=& V [\left<(\bar\psi D_0 \psi)^2\right>-\left<\bar\psi D_0 \psi\right>^2] \\
\chi_{e,i} &=& V [\left<P_i (\bar\psi D_0 \psi)\right>-\left<P_i\right>\left<\bar\psi D_0 \psi\right>], \quad i = \sigma , \tau \label{LATSUSC5}\\
\chi_{t,i} &=& V [\left<P_i (\bar\psi\psi)\right>-\left<P_i\right>\left<\bar\psi\psi\right>], \quad i = \sigma , \tau \label{LATSUSC6}\\\
\chi_{t,f} &=& V [\left<(\bar\psi\psi)(\bar\psi D_0 \psi)\right>-\left<(\bar\psi\psi)\right>\left<\bar\psi D_0\psi\right>] \label{LATSUSC7}
\end{eqnarray}
where $V=L_s^3 L_t$ is the volume; $D_0$ is the temporal component of the
Dirac operator $D$. 

\begin{figure*}[t!]
\includegraphics*[width=0.55\textwidth]{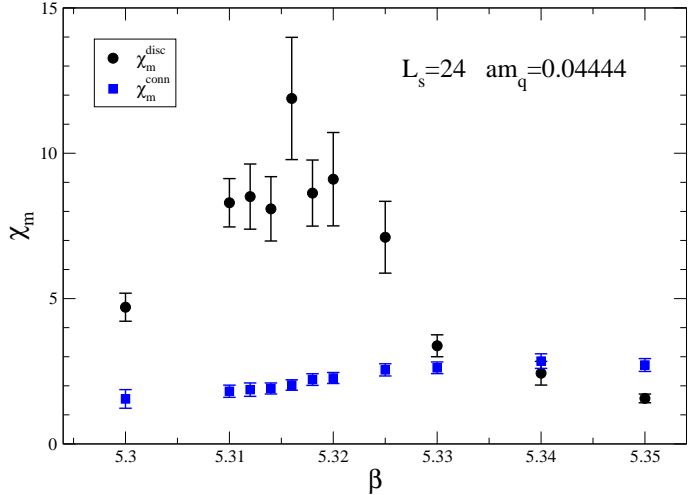}
\caption{Comparison between the connected and disconnectd component of $\chi_m$. The former is typically a small fraction of the connected component at the peak position.} \label{CHICONNFIG}
\end{figure*}

The connected component of the chiral susceptibility $\chi_m^{conn}$ has
not been measured for all of our simulations but only for a fraction
of them. The method used to extract $\chi_m^{conn}$ is the volume
source method without gauge fixing as described in Ref.\cite{jlqcd}.
The disconnected component gives the dominant contribution for large
volumes and small masses. The connected part is instead relevant at
small volumes and relatively large masses. For most of our lattices we
have determined the connected part only around the peak, and we have
considered it as a constant with respect to $\beta$. We estimate that
this is a good approximation within our errors.  A representative
example\footnote{For this lattice $\chi^{conn}_m$ was measured
for all points.} is shown in Fig.~\ref{CHICONNFIG} (taken at $L_s=24$,
$am_q=0.04444$). The typical contribution to $\chi_m$ of
$\chi_m^{conn}$ is less than about $15\%$ of $\chi_m^{disc}$ at the
peak value and is a slowly varying function of $\beta$.

A comprehensive list of measured values of $\langle P_\sigma\rangle$,
$\langle P_\tau\rangle$, $\langle\bar\psi\psi\rangle$,
$\langle\bar\psi D_0\psi\rangle$ and the lattice susceptibilities for
our MC simulation can be found in Appendix~\ref{appA}.

\section{Scaling analysis}\label{ANALYSIS}

The basic thermodynamic susceptibilities, i.e. the
specific heat $C_V$, the chiral susceptibility $\chi_m$ and the mixed
susceptibility $\chi_t$
\begin{eqnarray}
C_V &=& \frac{1}{VT^2} \frac{\partial^2}{\partial (1/T)^2} \ln Z \\
\chi_m &=& \frac{T}{V} \frac{\partial^2}{\partial m_q^2} \ln Z \label{chimeq}\\
\chi_t &=& \frac{T}{V} \frac{\partial^2}{\partial (1/T)\partial m_q} \ln Z
\end{eqnarray}
can be expressed as sums of the lattice susceptibilities
(\ref{LATSUSC1})-(\ref{LATSUSC7}) multiplied by regular functions of
$\beta$ and $am_q$. Specifically the specific heat $C_V$ is a function
of $\chi_{e,ij}$, $\chi_{e,i}$ and $\chi_{e,f}$; $\chi_t$ is a
function of $\chi_{t,i}$ and $\chi_{t,f}$. The contribution of other
susceptibilities entering in the expression of $C_V$ and $\chi_t$
involving the quark mass are expected to be negligible in the chiral
limit~\cite{Karsch:1986ec}.  As for $\chi_m$, $\partial/\partial
(m_q)$ in Eq.(\ref{chimeq}) is intended at constant temperature. Since
temperature depends not only on $\beta$ but also on $am_q$, the
physical $\chi_m$ is a combination of $\chi_m^{disc}$,
$\chi_m^{conn}$, $\chi_{t,\tau}$ and $\chi_{e,\tau\tau}$ with
computable coefficients.

By $C_V$ in the following we mean $\chi_{e,\sigma\sigma}$; the
analysis with $\chi_{e,i}$ and $\chi_{e,f}$ is similar and compatible
with respect to scaling.

We are interested in studying the singular behavior of $C_V$, $\chi_m$
and $\chi_t$ as the critical surface is approached which is given by
the most singular divergent quantity among the lattice
susceptibilities corresponding to a given termodynamical
susceptibility.

\subsection{Pseudocritical coupling}

One of the observables analyzed in the literature to understand the order
of the transition has been the position of the peaks of thermodynamic
susceptibilities as a function of $am_q$.  The position of all these
peaks happen to coincide at given $am_q$ and $L_s$, thus defining a
unique (pseudo)critical coupling $\beta_c(am_q)$. Previous works in the
literature assume $\tau\propto \beta_0 - \beta$, a choice usually
based on a strict analogy between QCD and the $O(4)$ statistical
model. In fact the correct thermodynamical reduced temperature is
given by Eq.~(\ref{taudef}).  In principle the dependence of
$a(\beta,am_q)$ on $am_q$ could be measured by use of independent
quantities (see e.g.~\cite{massapi}). We will try a fit of the
position of $\beta_c$ by a form like Eq.~(\ref{pcbscal}) or
(\ref{pcbscal1}), which is expected to be valid at sufficiently small
values of $am_q$. To extend the range of validity of the approximation
the quadratic terms proportional to $am_q^2$, $am_q(\beta_0-\beta)$
and $(\beta_0-\beta)^2$ may be added:
\beq
\tau \propto (\beta_0 - \beta )+ k_m am_q + k_{m^2} (am_q)^2 + k_{m\beta} am_q (\beta_0 - \beta ) \, .
\label{taudef2}
\eeq
A term $k_{\beta^2}(\beta_0 - \beta )^2$ turns out to be
negligible. In fact one can write the lattice spacing $a$ as:
\beq
a = \frac{1}{\Lambda(\beta,am_q)} \left(\frac{\beta}{4N_c b_0}\right)^{\frac{b_1}{2b_0^2}}\exp\left(-\frac{\beta}{4N_c b_0}\right)
\eeq
where the deviation from asymptotic scaling are represented by the
fact that the term $\Lambda(\beta,am_q)$ is $\beta$ dependent. This
term is slowly varying with $\beta$ and can be well described by a
linear function of $\beta$ with coefficients depending on $am_q$ in
the relevant range of $\beta$'s. The $k_{\beta^2}$ coefficient is
given by $\partial^2 \ln a/\partial\beta^2|_{\beta=\beta_0,am_q=0}$
can be computed and can be neglected within errors. The other unknown
parameters can be fitted to the data.

\begin{figure*}[t!]
\includegraphics*[width=0.95\textwidth]{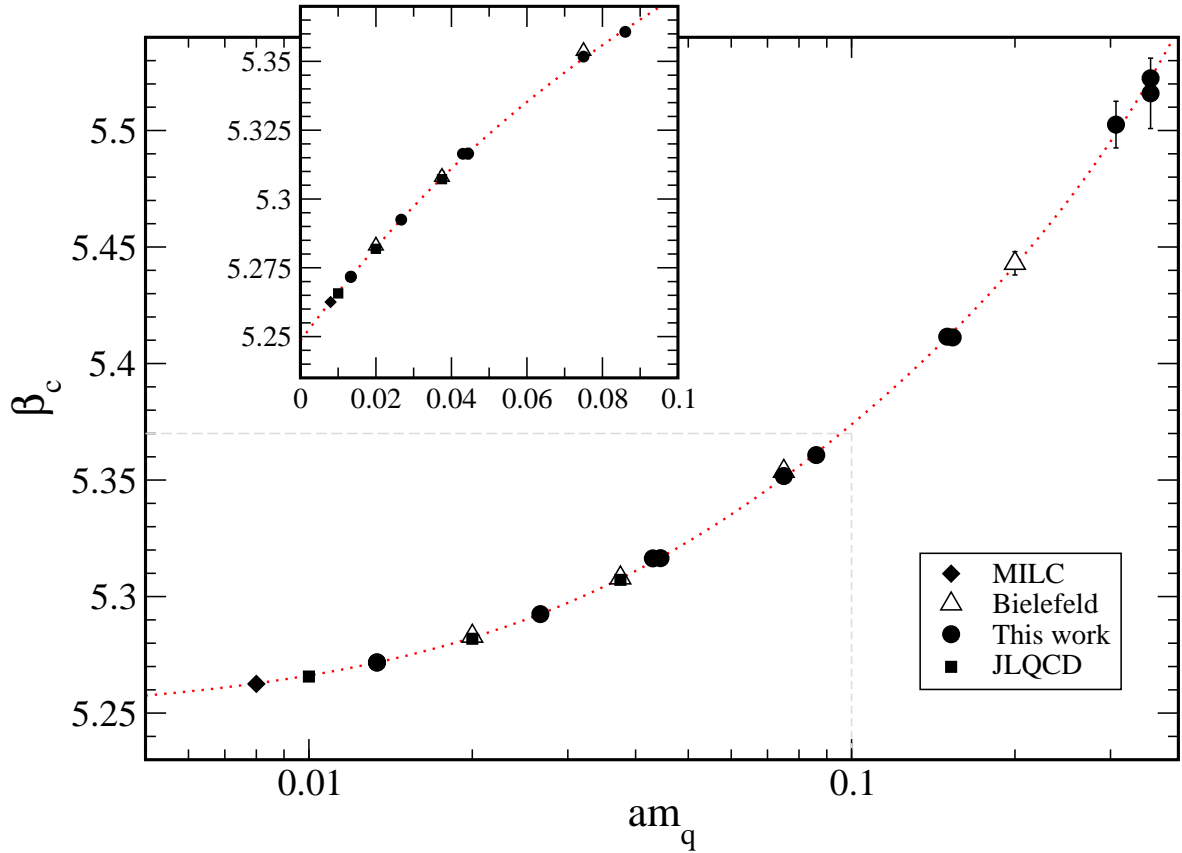}
\caption{(Pseudo)critical couplings determined in this work are shown together with other determinations form the literature. Logarithmic scale is used for the horizontal scale in the main figure, while a linear scale is used for the smaller inset figure. The dotted line is the best fit ($\chi^2/d.o.f=0.89$, $d.o.f.=15$) curve including $am_q^2$ and $am_q(\beta_0-\beta_c)$ terms for masses $am_q<0.4$.}\label{PCC}
\end{figure*}
Fig.~\ref{PCC} shows the critical line $\beta_c(am_q)$. Our
determinations are reported together with a collection of world
data\footnote{Data of the JLQCD collaboration are taken from
Ref.~\cite{jlqcd}. We thank E. Laermann and C. DeTar for providing us
with the data of the Bielefeld group and of the MILC collaboration
respectively.}. A good agreement among different determinations can be
appreciated.

The expected variation of $\tau$ as a function of $am_q$ or $L_s$ is
given by Eq.~(\ref{pcscale}) or (\ref{pcscale1}):
\begin{eqnarray}
\tau &=& k_{\tau} (am_q)^{1/\nu y_h} \,\,\,\,\,\,\, \rm{or} \label{tau1}\\
\tau &=& k'_{\tau} L_s^{1/\nu} \,\,\,. \label{tau2}
\end{eqnarray}

Notice that for a first order transition the exponent of $am_q$ is $1$
and the term on the right hand side of Eq.~(\ref{tau1}) can be
reassorbed in $k_m$ so that such term can be discriminated only for a
second order scaling behavior.

The unknown coefficients $k_m$, $k_{m^2}$, $k_{m\beta}$ and $k_{\tau}$
of the expansion of $\tau$ around the critical point are determined by
use of a best fit procedure. We start from the form
Eq.~(\ref{tau2}). In agreement with previous works we find that the
position of the peaks does not depend on the lattice size
\cite{karsch1,jlqcd}, i.e. that $k'_\tau = 0$ within errors implying
that no information can be obtained about the order of the transition
(first order or second order $O(4)$ or $O(2)$).  The quality of the
fit assuming $k'_\tau =0$ is shown in Fig.~\ref{PCC} and the resulting
coefficients are listed in the first line of
Table~\ref{bcfittable}. These coefficients are obtained by a best fit
up to a maximum value of $am_q$, $(am_q)_{max}$, which is then
extrapolated to zero. They are stable and consistent with a linear fit
at low values of $am_q$ ($<0.0267$) as shown in the second line of
Table~\ref{bcfittable}.

The scaling of Eq.~(\ref{pcscale1}) or (\ref{tau2}) assumes that $aL_s
m_\pi \gg 1$ but the correlation length $\xi$ can be comparable with
$L_s$, which is certainly true sufficiently close to the critical
point in case of a second order or weak first order chiral transition.

We have then analyzed the scaling of the form Eq.~(\ref{pcscale}) or
(\ref{tau1}). If the transition is first order the analysis coincides
with the analysis done for the scaling Eq.~(\ref{tau2}) and $k_m$ is
in fact $k_m-k_\tau$.

If the transition is $O(4)$ a similar analysis can be performed
(similar results hold for $O(2)$ or mean field). The $\chi^2/d.o.f$ is
acceptable, the result for the coefficients is shown in the third line
of Table~\ref{bcfittable}. $k_\tau$ is consistent with zero, and the
result is therefore compatible with that of the first line. However
the fit becomes unstable if we try to extrapolate to low masses
keeping only the linear term of Eq.(\ref{taudef2}) (line 4 of
Table~\ref{bcfittable}).

The critical coupling $\beta_0$ and the coefficient $k_m$ are stable
both for first order and $O(4)$ behavior and can thus be confidently
estimated. Our final estimates, obtained by a weighted average of
linear and quadratic fits, are $\beta_0=5.2484(5)$ and $k_m=1.82(8)$
for a first order transition; $\beta_0=5.2435(25)$ and $k_m=1.13(19)$
for $O(4)$. Other terms cannot be reliably estimated with present
data. In particular we are not able to discriminate the contribution
of quadratic terms from that coming from $k_{\tau}$ and consequently
it is not possible to establish the order of the transition by looking
at the (pseudo)critical couplings alone.

\begin{table}[b!]
\caption{Best fit parameters for the scaling of the pseudocritical coupling. Different kind of fits are explained in the text. $k'_{\tau}$ is set to zero as explained in the text.}\label{bcfittable}
\begin{tabular}{|c|c|c|c|c|}
\hline
$\beta_0$ & $k_m$ & $k_{\tau}$ & $k_{m^2}$ & $k_{m\beta}$  \\
\hline
\hline 5.2484(4) & 1.84(7) & $\equiv 0$ & -0.3(2.4) & -4.3(2.7)\\
\hline 5.2481(13) & 1.75(13) & $\equiv 0$ & $\equiv 0$ & $\equiv 0$ \\
\hline 5.2430(40) & 1.20(60) & -0.1(1) & -7(6) & 6(6)\\
\hline 5.2437(21) & 1.12(16) & -0.134(46) & $\equiv 0$ & $\equiv 0$ \\
\hline 
\end{tabular}
\end{table}

\begin{figure*}[t!]
\includegraphics*[width=0.95\textwidth]{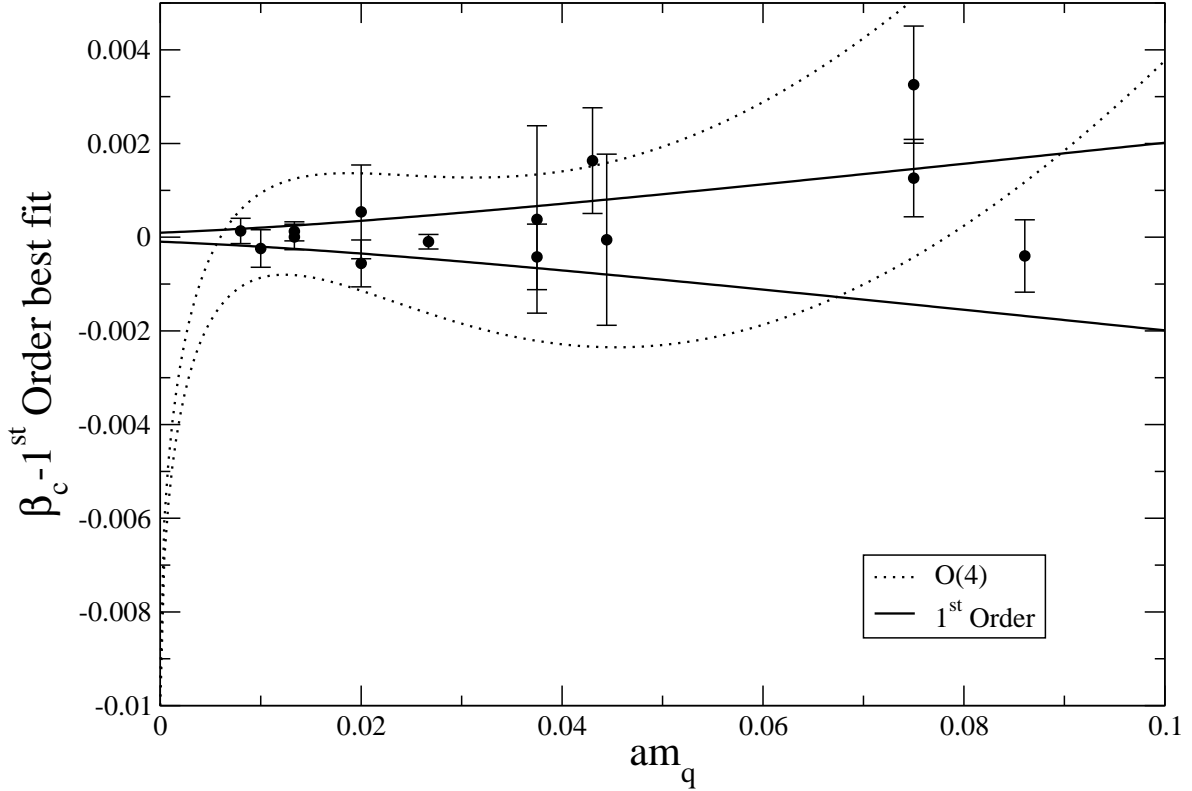}
\caption{Different behavior of predicted (pseudo)critical couplings for a first order transition and a second order $O(4)$. Present data do not permit a clear discrimination using only $\beta_c$. The 1$\sigma$ band is displayed. Continuous lines correspond to a first order transition while the dotted lines are the prediction for the $O(4)$ symmetry.}\label{PCC2}
\end{figure*}

The possibility to discriminate between the first order and $O(4)$
behavior from the measurement of $\beta_c$ is in practice very
faint. Fig.~(\ref{PCC2}) shows the different predictions for the
(pseudo)critical coupling based on available data. If we discard the
high values of the masses a possible difference would only be visible
at very small bare quark masses. Using estimates of $k_{\tau}$ shown
in Table~\ref{bcfittable} a quark mass of about $0.003$ should be
used, which requires a big numerical effort.

We would like to remark that the explicit dependence of $\tau$ on
$am_q$ is in any case necessary to fit present data. Fitting to a
function of the form
\begin{equation}
\beta_c = \beta_0 + c_\tau (am_q)^{y_t/y_h}
\end{equation}
with $O(4)$ values for the exponents, gives a $\chi^2/d.o.f\simeq 21$
in the interval up to $(am_q)_{max}=0.075$, which decreases as
$(am_q)_{max}$ decreases and is $\approx 2$ at $(am_q)_{max}=0.02$.

As a final remark, the dependence on $\beta$ and $am_q$ of the lattice
spacing $a$ Eq.~\ref{taudef} can be measured from other observables
(see e.g.~\cite{massapi}). In particular our estimate for $k_m$ for a
first order transition (line 1 of Table~\ref{bcfittable}) is
compatible with those of Ref.~\cite{massapi} $k_m\approx 1.95$
(affected by errors of order 20\%). We notice that $k_\tau=0$ implies
that $\tau=0$ on the critical line, or, by Eq.~(\ref{spacingexp}) that
the critical temperature is independent of $am_q$ near the chiral
point.

\subsection{Scaling at fixed $am_q L_s^{y_h}$}

As explained in details in Section~\ref{STRATEGY}, we have adopted a
novel strategy in order to simplify the two scales problem. We assume
$O(4)$ --- or $O(2)$ --- critical behaviour and we use this assumption
to fix a dependence between $am_q$ and $L_s$ in our Run1 and Run2 so
as to fix the second scaling variable in Eq.~(\ref{scal2}) and reduce
the problem to a one scale problem: in this case the only assumption
is $O(4)$ itself. This allows us to test whether $O(4)$ is consistent
or not with data without any further approximation.

We fixed $am_q L_s^{y_h}=const$ with $y_h=2.49$ that is the value
expected for $O(4)$ and $O(2)$ critical behavior, with $const=74.7$ for
our Run1 and $const=149.4$ for Run2. The following scaling formulas
should hold (see Eq.s~(\ref{scalcal})~and~(\ref{scalord})):
\begin{eqnarray}
C_V(\tau, L_s) - C_0 = L_s^{\alpha/\nu} \Phi_C(\tau L_s^{1/\nu})\label{CV}\\
\chi_m(\tau, L_s) = L_s^{\gamma/\nu} \Phi_\chi(\tau L_s^{1/\nu})\label{CHI}
\end{eqnarray}
and in particular the peaks of the specific heat and $\chi_m$ should
scale as Eq.s~(\ref{scalmax}).  

The subtraction of the non critical part $C_0$ for the specific heat
is needed. In principle it can be obtained as a parameter from the fit
to the maxima of the specific heat. However since we also have data at
$\beta$'s different from the pseudocritical coupling, we were able to
perform a direct measurement of this quantity. Appendix~\ref{appB}
reports the details of the study of the background for $C_V$. Our
final estimate for the background is $C_0 (\beta)= 0.400(43) -
0.0663(83) \beta$. Note that the $\beta$ dependence is very weak and
assuming a constant value for the background $C_0$ does not modify the
following analysis. No dependence of $C_0$ on $am_q$ is observed.

For the susceptibility of the chiral condensate $\chi_m$, for the
moment we do not operate any bakground subtraction.

\begin{figure*}[t!]
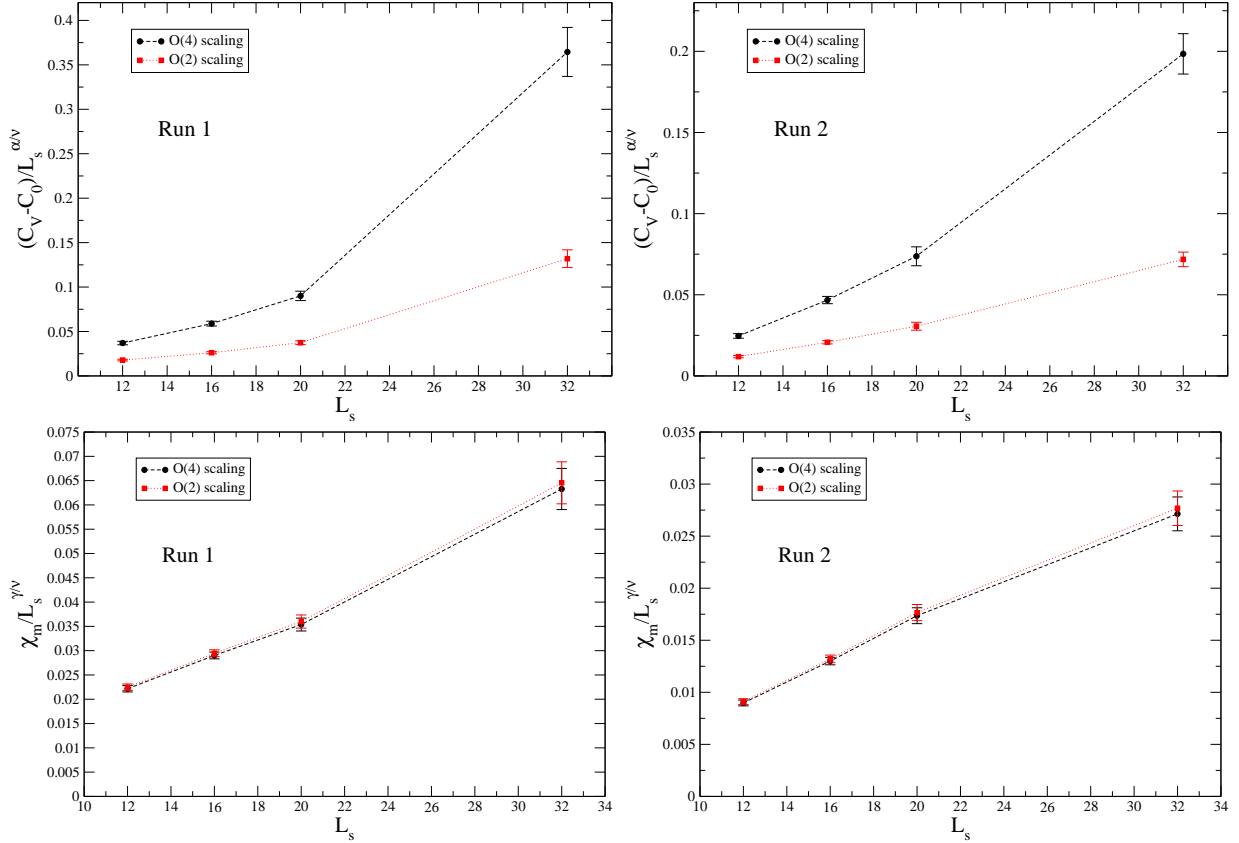

\includegraphics*[width=0.49\textwidth]{Cv_max_Run1.eps}
\includegraphics*[width=0.49\textwidth]{Cv_max_Run2.eps}\\
\includegraphics*[width=0.49\textwidth]{Chi_max_Run1.eps}
\includegraphics*[width=0.49\textwidth]{Chi_max_Run2.eps}
\caption{Specific heat (top) and $\chi_m$ (bottom) peak value for Run1 (left) and for Run2 (right), divided by the appropriate powers of $L_s$ (Eq.s~\ref{CV}-\ref{CHI}) to give a constant. Both the $O(4)$ and $O(2)$ critical behaviors are displayed. Notice that for the case of $\chi_m$ the ratio $\gamma/\nu$ have almost the same numerical value so that the two curves are almost indistinguishable.}\label{R12}
\end{figure*}

The measured peak values for the subtracted specific heat $C_V-C_0$
and chiral condensate susceptibility $\chi_m$ for Run1 and Run2 are
shown in Fig.~\ref{R12}. They are evaluated on the curve obtained by
reweighting.
The figure shows the peak values of susceptibilities rescaled by the
appropriate power of the spatial lattice size Eq.~(\ref{scalmax}). If
the scaling laws (\ref{scalmax}) would hold, the displayed quantity
should be a constant. Visibly this is not the case.
\begin{figure*}[bt!]
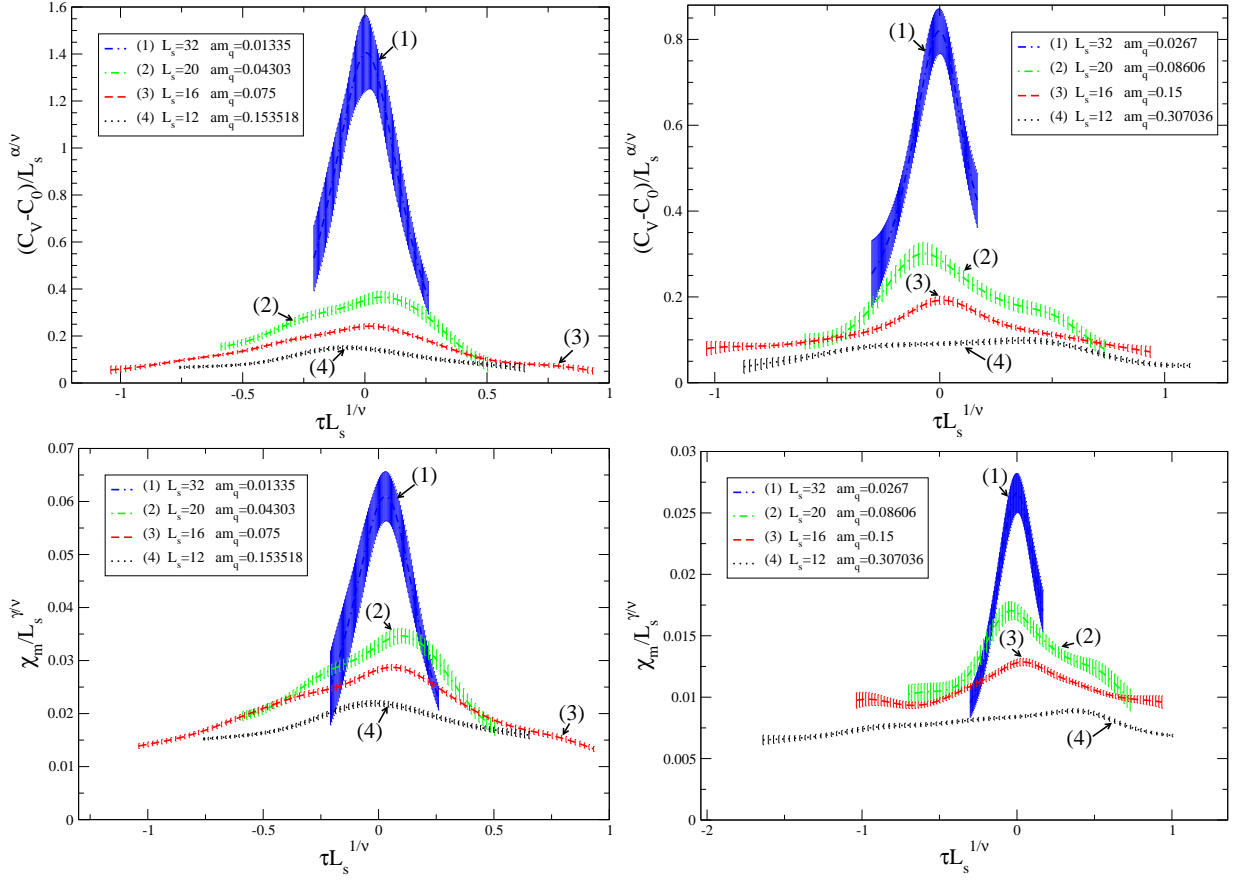

\includegraphics*[width=0.49\textwidth]{Cv_Run1.eps}
\includegraphics*[width=0.49\textwidth]{Cv_Run2.eps}\\
\includegraphics*[width=0.49\textwidth]{Chi_Run1.eps}
\includegraphics*[width=0.49\textwidth]{Chi_Run2.eps}
\caption{Scaling of the specific heat (top) and $\chi_m$ (bottom) for Run1 (left) and for Run2 (right), see Eq.s~(\ref{scalcal}) and (\ref{scalord}). The curves are obtained by reweighting.}\label{R12SCA}
\end{figure*}

The $O(4)$ and $O(2)$ critical behavior is clearly in contradiction
with the lattice observation.  In particular $O(4)$ and $O(2)$ scaling
predicts no singular behavior in the $L_s\rightarrow\infty$ limit for
the specific heat as the critical exponent $\alpha$ is negative.  This
means that as $L_s$ is increased, the singular part of $C_V$ should
decrease with volume, i.e. that the specific heat should not grow
which is in clear contrast with the data. Also for the chiral
condensate susceptibility $\chi_m$ the predicted exponents fail to
reproduce lattice data.  In either case the $\chi^2/d.o.f.$ of the fit
with a constant function excludes the behavior of scaling law
(\ref{scalmax}).  The full scaling laws
Eq.s~(\ref{scalcal})~and~(\ref{scalord}) were also studied
(see~Fig.(\ref{R12SCA})). The horizontal scale was obtained by fitting
the pseudocritical temperature as described above. As one
can expect from the previous discussion, data don't scale according to
the predicted laws.

Similar figures are obtained assuming $O(2)$ symmetry.

\subsection{Scaling at $L_s\rightarrow\infty$}

\begin{figure*}[tb!]
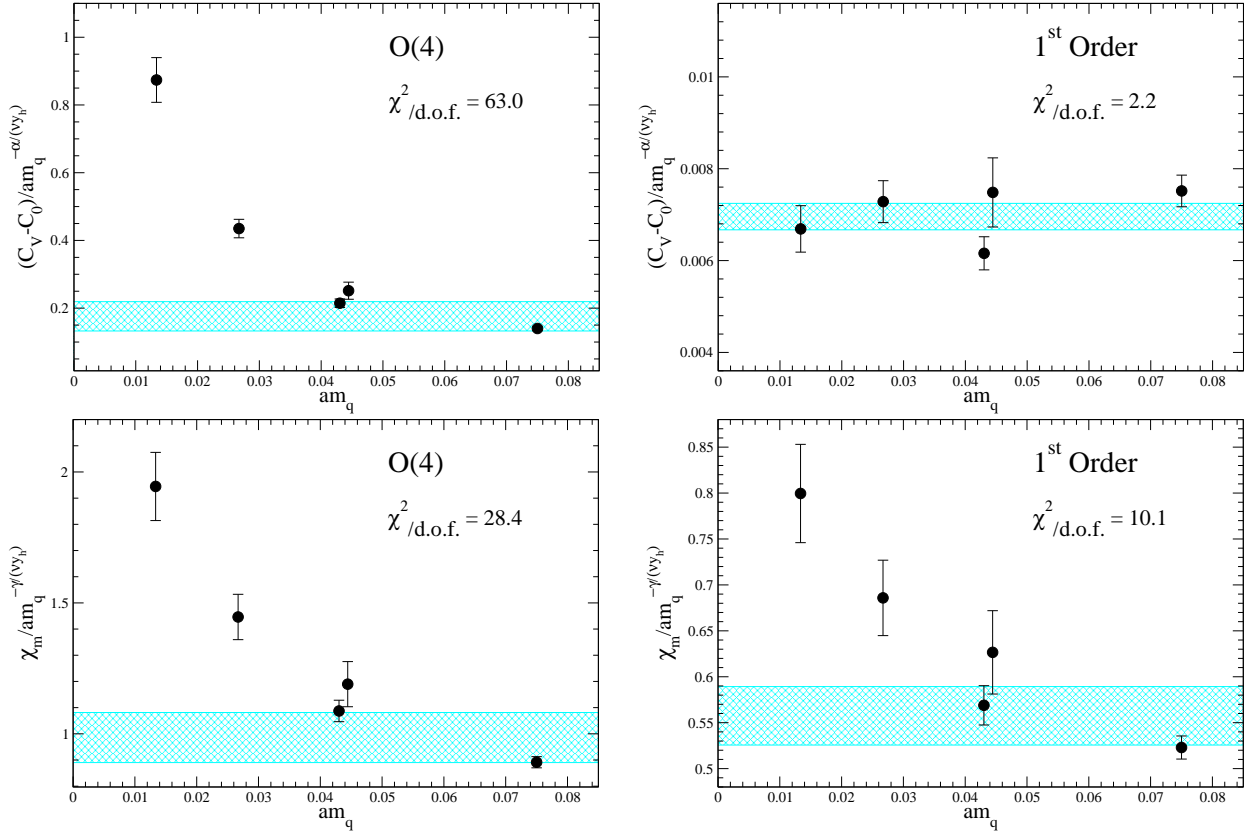

\includegraphics*[height=5.5cm]{Cv_max_O4.eps}\hfill
\includegraphics*[height=5.5cm]{Cv_max_1st.eps}\\
\includegraphics*[height=5.5cm]{Chi_max_O4.eps}\hfill
\includegraphics*[height=5.5cm]{Chi_max_1st.eps}
\caption{Specific heat (top) and $\chi_m$ (bottom) peak scaling for $O(4)$ (left) and first order (right).}\label{SHCHI}
\end{figure*}

\begin{figure*}[htb!]
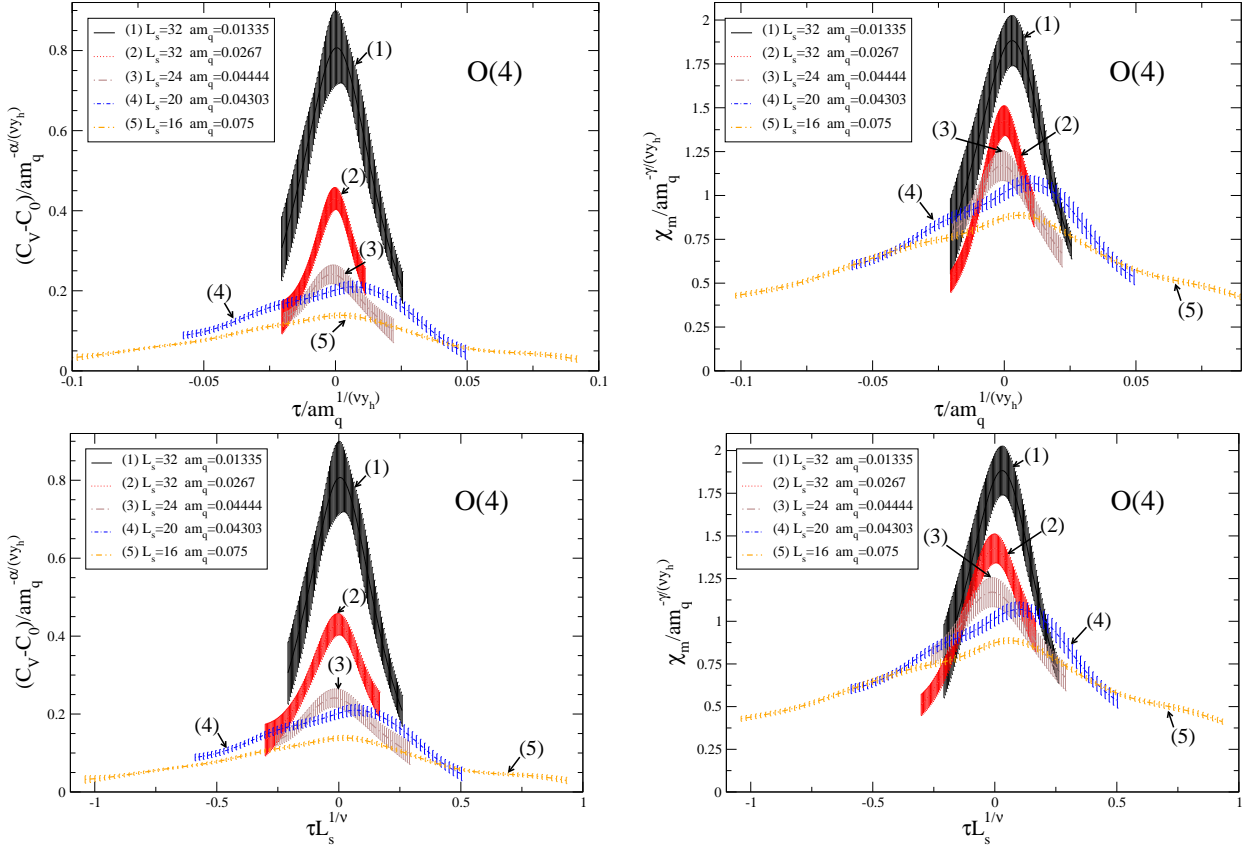

\includegraphics*[height=5.7cm]{Cv_O4.eps}\hfill
\includegraphics*[height=5.7cm]{Chi_O4.eps} \\
\includegraphics*[height=5.5cm]{Cv_O4-Ls.eps}\hfill
\includegraphics*[height=5.5cm]{Chi_O4-Ls.eps} \\
\caption{Comparison of specific heat (left) and $\chi_m$ (right) scaling for $O(4)$. Eq.\ref{scalcal2}, \ref{scalord2} (top) and  Eq.\ref{scalcal}, \ref{scalord} (bottom).}\label{SHCHISCA}
\end{figure*}

A further scaling test can be done supposing that the lattice size is
much larger that all other relevant physical lengths. In such a case
one expects that the system show the same behavior of an infinite
system. The scaling laws expected in this case are those of
Eq.s~(\ref{scalcal1})~and~(\ref{scalord1}). These equations
predict no dependence on the parameter $L_s$. This is the same
assumption used in previous scaling analyses of the chiral
transition~\cite{karsch2,jlqcd,milc}.  

The alternative possibility, illustrated in section~\ref{STRATEGY}, is
to keep $\tau L_s^{1/\nu}$ (i.e. $\xi/(aL_s)$) fixed thus remaining
with the scaling equations Eq.s~(\ref{scalcal2})~and~(\ref{scalord2}).
Physically this means that the correlation length $\xi$ is not small
compared to $L_s$, which is certainly true in the vicinity of the
critical point.

These scaling laws are only expected to hold for large values of $am_q
L_s^{y_h}$ and small masses.

The difference between the two alternatives is only visible by
considering the width of the susceptibilities peaks, the heights
having the same behavior (see Eq.s(\ref{scalmax1})).

We have thus first performed the analysis of the scaling of the maxima
of the specific heat $C_V$ and $\chi_m$.

\begin{figure*}[htb!]
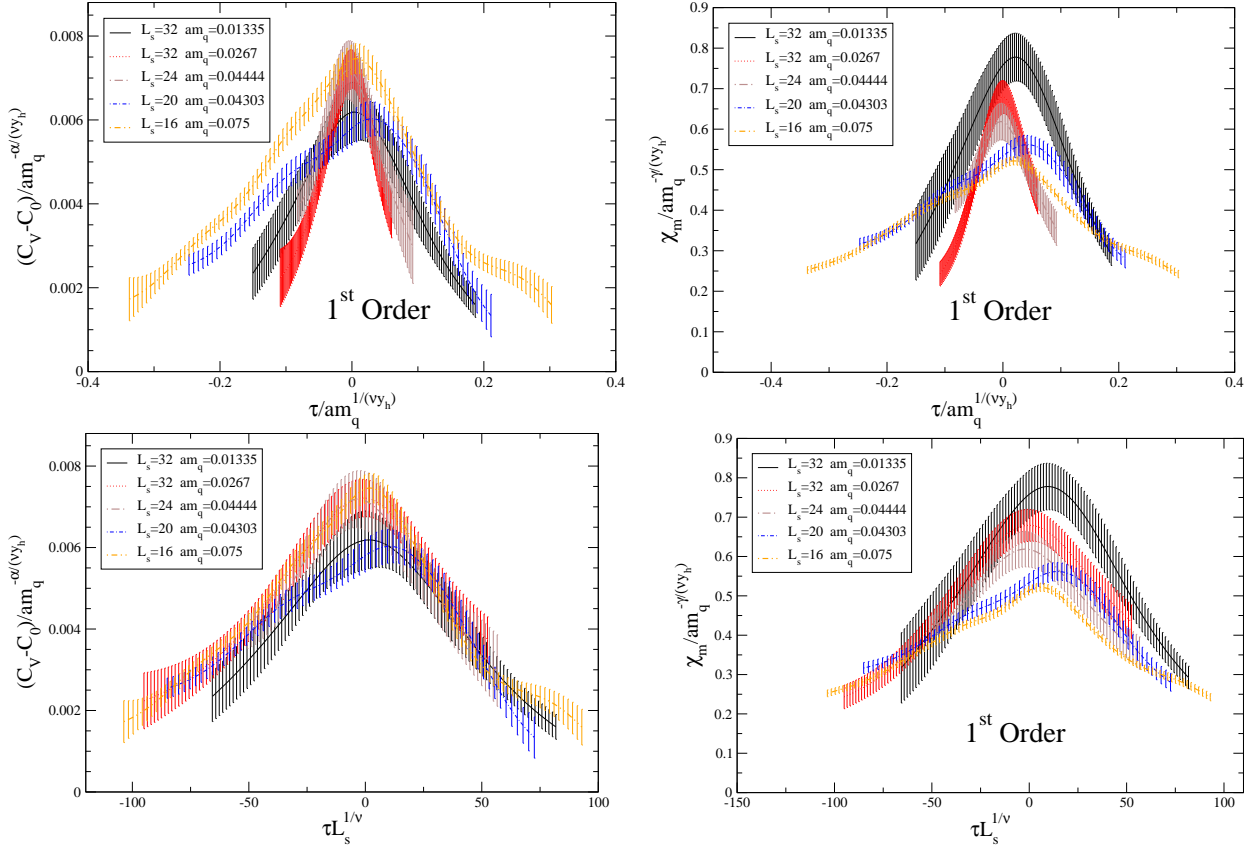

\includegraphics*[height=5.7cm]{Cv_1st-m.eps}\hfill
\includegraphics*[height=5.7cm]{Chi_1st-m.eps} \\
\includegraphics*[height=5.5cm]{Cv_1st.eps}\hfill
\includegraphics*[height=5.5cm]{Chi_1st.eps} \\
\caption{Comparison of specific heat (top) and $\chi_m$ (bottom) scaling for first order. Eq.\ref{scalcal2} and \ref{scalord2}.}\label{SHCHISCA2}
\end{figure*}

We have tested the different second order critical behaviors
compatible with the scenario of Ref.~\cite{wilcz1}, namely $O(4)$,
$O(2)$, Mean Field and first order. The peak value of the specific
heat and of $\chi_m$ divided by the appropriate power of the quark
mass should be a constant. These ratios are shown in Fig.(\ref{SHCHI})
both for $O(4)$ and first order. The figure shows also the confidence
region from a fit with a constant value together with the
corresponding $\chi^2$/d.o.f.

From the values of the $\chi^2$/d.o.f. it is easily seen that the
second order critical behavior is not compatible with data.  It must
be noticed that, although the validity region of the scaling law is
not known \textit{a priori}, so that the upper mass limit for the fits
is somewhat arbitrary, if we further restrict the mass region, the
$\chi^2/d.o.f.$ for the $C_V$ fits tend to increase.  We stress that
also in previous studies~\cite{karsch2,jlqcd,milc} the values found
for the susceptibility peaks were not compatible with the critical
indexes of $O(4)$, $O(2)$ and Mean Field.  On the contrary the first
order behavior looks compatible with data (even if the $\chi^2/d.o.f.$
is $\sim 2$) for the specific heat. We will discuss the scaling of
$\chi_m$ in Sect.\ref{mag}.

The scaling of susceptibilities at all $\beta$ values can also be
investigated. The situation is depicted in Fig.~\ref{SHCHISCA} for
$O(4)$: all the curves should coincide within errors if there were
$O(4)$ scaling. Similar features are observed for $O(2)$ and Mean
Field.  No scaling is observed: this is clearly the case for the
maxima of the susceptibilities as discussed above, but it is also true
for the width of these curves.  The analogous figure using first order
indexes is shown in Fig.~\ref{SHCHISCA2}. Scaling is observed for the
specific heat in the form Eq.s~(\ref{scalcal2}) and (\ref{scalord2})
and not in the form (\ref{scalcal1}) and (\ref{scalord1}), which does
not describe the widths of the peaks. Again for $\chi_m$ we postpone
the discussion to Sect.\ref{mag}.

\begin{figure*}[th!]
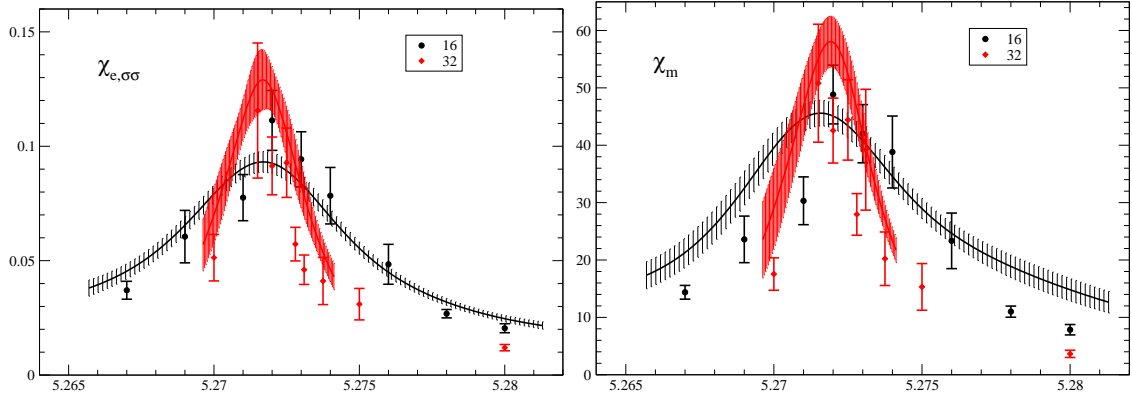

\includegraphics*[width=0.45\textwidth]{pssusc1632.eps}
\includegraphics*[width=0.45\textwidth]{chisusc1632.eps}
\caption{Comparison between $\chi_{e,\sigma\sigma}$ (left) and $\chi_m$ (right) for $am_q=0.01335$ and $L_s=16,32$.}\label{pschi1632}
\end{figure*}
Runs with higher values of the masses ($am_q>0.075$) do not obey the
scaling laws. The one with $L_s=16$ and $am_q=0.01335$, which should
coincide with the $L_s=32$ at the same bare mass, in case of scaling
Eq.s~(\ref{scalcal1}) and (\ref{scalord1}) is instead different (see
Fig.~\ref{pschi1632}). For scaling Eq.s~(\ref{scalcal2}) and
(\ref{scalord2}), the maximum should be the same, and the widths
should differ by a factor of 8, and this is not the case. We have
carefully checked the stability of the curves obtained by reweighting
against variations of the statistics, e.g. by discarding single data
points from the analysis. This lack of scaling can be interpreted as
due to the small value (4.5) of the parameter $aL_sm_\pi$,
invalidating the limit bringing to Eq.s~(\ref{scalcal1}),
(\ref{scalord1}), (\ref{scalcal2}) and (\ref{scalord2}). A similar
effect could be responsible for the observed increase of the peak with
the volume observed in previous studies~\cite{jlqcd}.

A more careful study of this effect will be done in the future.

\subsection{Magnetic equation of state}\label{mag}

\begin{figure*}[th!]
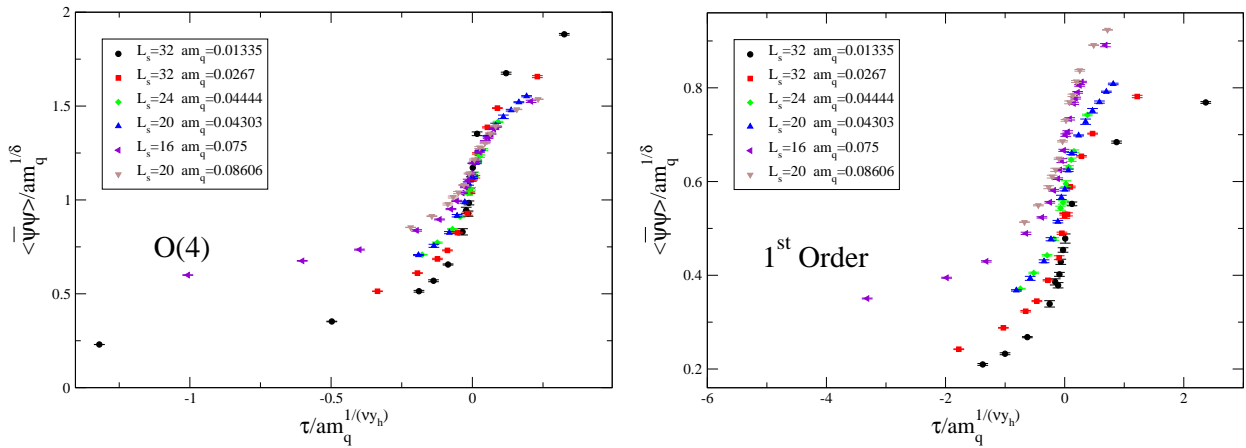

\includegraphics*[width=0.49\textwidth]{eqstO4-nosub.eps}\hfill
\includegraphics*[width=0.49\textwidth]{eqst1st-nosub.eps}
\caption{Magnetic equation of state using non subtracted data for $O(4)$ (left) and first order (right).}\label{noscal}
\end{figure*}

As a further test of scaling we check the equation of state
Eq.~(\ref{eqstate}). No scaling whatsoever is observed, neither
$O(4)$-$O(2)$ nor first order, if the raw measured data are introduced
in Eq.(\ref{eqstate}) (see Fig.\ref{noscal}).
Indeed as $\langle\bar\psi\psi\rangle$ is different from 0 at large
$\beta$ a subtraction is needed: the critical part of the chiral
condensate has to be zero far above the critical region. A tentative
way to understand the non critical background can be to assume it
equal to the value $\langle\bar\psi\psi\rangle_\infty$ of
$\langle\bar\psi\psi\rangle$ at $\beta=\infty$. This can be computed
analytically and numerically on the flat configuration
$U_{\mu\nu}=1$. The result is
\begin{equation}
\langle\bar\psi\psi\rangle_\infty (am_q)= \frac{3}{L_s^3 L_t}\sum_{k_1,k_2,k_3,k_4} \frac{am_q}{(am_q)^2+\sum_{i=1,2,3}(\sin\frac{\pi}{L_s k_i})^2+(\sin\frac{\pi}{L_t (k_4+1/2)})^2}
\end{equation}
which is almost a linear function in the mass range $am_q<0.1$.

\begin{figure*}[b!]
\includegraphics*[width=0.95\textwidth]{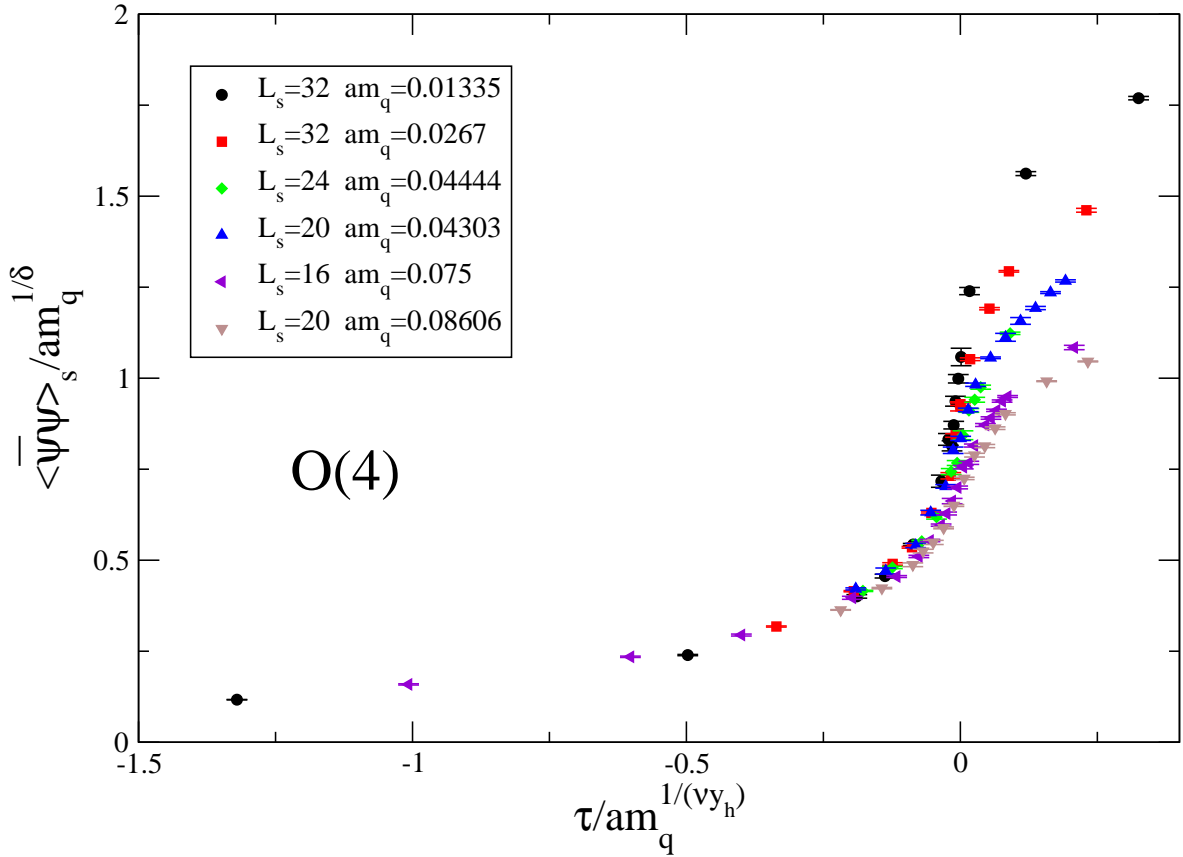}
\caption{Equation of state for $O(4)$, obtained by subtraction of $\langle\bar\psi\psi\rangle$ at $\beta=\infty$.}\label{eqstfig}
\end{figure*}

We then plot the subtracted value
$\langle\bar\psi\psi\rangle_s\equiv\langle\bar\psi\psi\rangle-\langle\bar\psi\psi\rangle_\infty$
rescaled as $\langle\bar\psi\psi\rangle_s/(am_q)^{1/\delta}$ versus
$\tau (am_q)^{-1/(\nu y_h)}$ to test the scaling
Eq.~(\ref{eqstate}). Fig.~\ref{eqstfig} shows the result for $O(4)$
(similar figures being obtained with $O(2)$ and mean field). Visibly
$O(4)$ scaling is not obeyed.  An analogous investigation has been
performed in Ref.~\cite{milc}, without the subtraction of the
$\beta=\infty$ term: also in that case results were in disagreement
with $O(4)$ scaling.

\begin{figure*}[tb!]
\includegraphics*[width=0.95\textwidth]{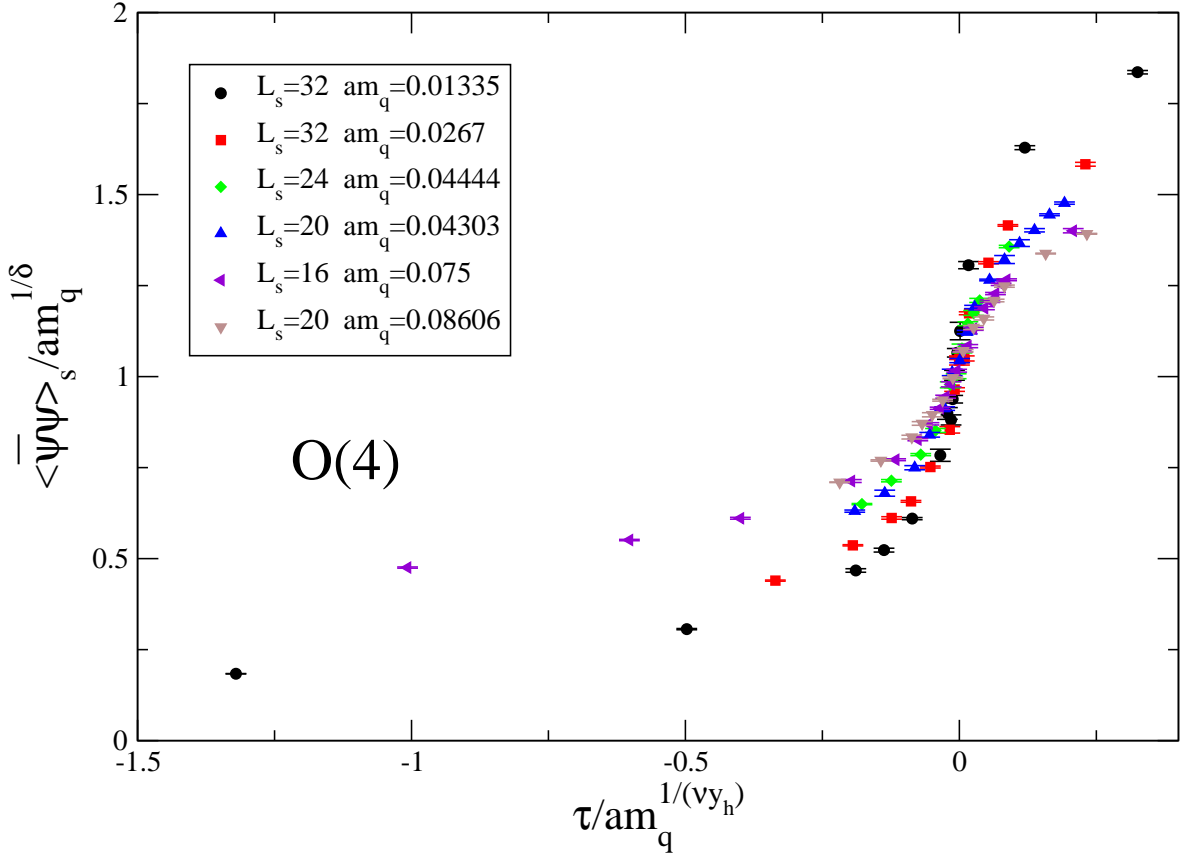}
\caption{Equation of state for $O(4)$, obtained by imposing scaling at $\tau=0$.}\label{eqstfig3}
\end{figure*}

An alternative procedure is to subtract for each $(am_q,L_s)$ the
value at the largest measured $\beta$. The result is consistent with
Fig.~\ref{eqstfig}. A third possibility is to impose scaling at
$\tau=0$ and look if it is obeyed at $\tau\neq 0$. The result is shown
in Fig.~\ref{eqstfig3} and again there is no scaling.

If the analysis of \cite{wilcz1} is correct, our results, which exclude
a second order critical behavior, should then imply a first order
chiral transition.

\begin{figure*}[b!]
\includegraphics*[width=0.95\textwidth]{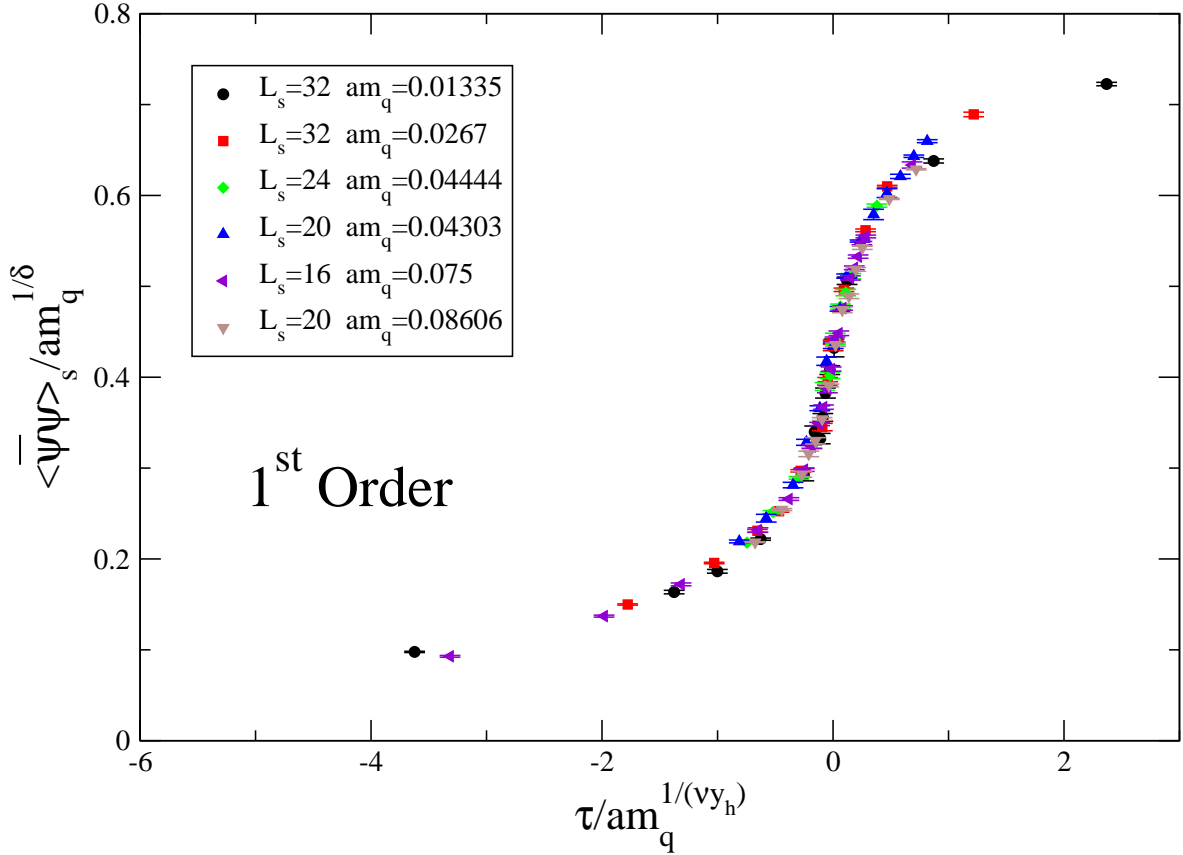}
\caption{Equation of state for first order.}\label{eqstfig2}
\end{figure*}

As a test of this possibility we have repeated the scaling analysis of
the equation of state using first order critical indexes: the three
procedures described above are consistent with each other and the
result is shown in Fig.~\ref{eqstfig2}.
A good scaling is observed.

We can investigate the consequence of the subtraction needed to
isolate the critical part of $\langle\bar\psi\psi\rangle$ on the
scaling of $\chi_m$. If
$\langle\bar\psi\psi\rangle-\langle\bar\psi\psi\rangle_\infty =
am_q^{1/\delta} F(\tau/am_q^{1/\nu y_h})$ by differentiating with
respect to $am_q$ at fixed temperature we get
\begin{equation}
\chi_m-\frac{\partial}{\partial am_q}\langle\bar\psi\psi\rangle_\infty = \frac{1}{\delta} am_q^{1/\delta-1} F - \frac{1}{\nu y_h} am_q^{1/\delta-1} (\tau/am_q^{1/\nu y_h}) F'
\end{equation}
Keeping in mind $1/\delta=(d-y_h)/y_h$, $\gamma=(2y_h-d)/y_t$ and $\nu=1/y_t$ we find, at $\tau=0$
\begin{equation}
am_q^{\gamma/\nu y_h} \left( \chi_m - \frac{\partial}{\partial am_q}\langle\bar\psi\psi\rangle_\infty \right) = \frac{1}{\delta} F(0)\label{eqst_fondo}
\end{equation}
The quantity which scales is not $am_q^{\gamma/\nu y_h} \chi_m$ but a
term $am_q^{\gamma/\nu y_h}
\partial\langle\bar\psi\psi\rangle_\infty/\partial (am_q)$ must be added to
it to get scaling. The subtraction of $\langle\bar\psi\psi\rangle$,
due to the explicit breaking of chiral symmetry at $am_q\neq 0$,
implies a subtraction for $\chi_m$ (which in the mass range of
interest is almost constant). 

This suggests to repeat the test of scaling for $\chi_m$
(Fig.s~\ref{R12},\ref{R12SCA},\ref{SHCHI} and \ref{SHCHISCA}) by
introducing a subtraction by a constant to be determined. The content
of Fig.~\ref{pschi1632} instead stays unchanged since the curves refer
to the same value of the mass.

\begin{figure*}[tb!]
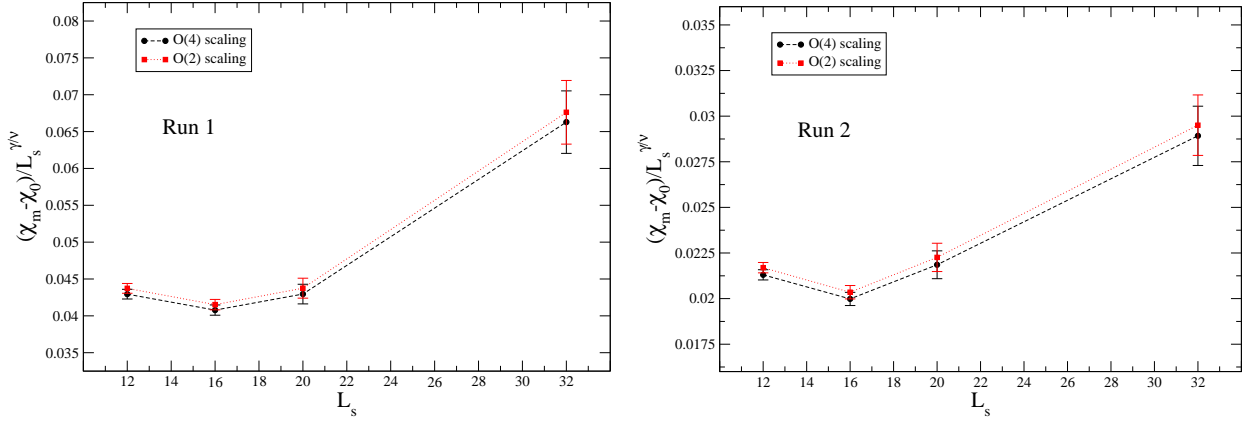

\includegraphics*[width=0.49\textwidth]{Chi_max_Run1_2.eps}\hfill
\includegraphics*[width=0.49\textwidth]{Chi_max_Run2_2.eps}
{%
\renewcommand{\thefigure}{\ref{R12}$'$}%
\caption{\label{R12p}Scaling of the maxima of $\chi_m-\chi_0$ at fixed $am_qL_s^{y_h}$ for Run1 (left) and Run2 (right). Two curves are shown corresponding to $O(4)$ (circles) and $O(2)$ (squares) critical behavior. The value of $\chi_0$ is obtained by a best fit procedure.}%
}
\end{figure*}

\begin{figure*}[tb!]
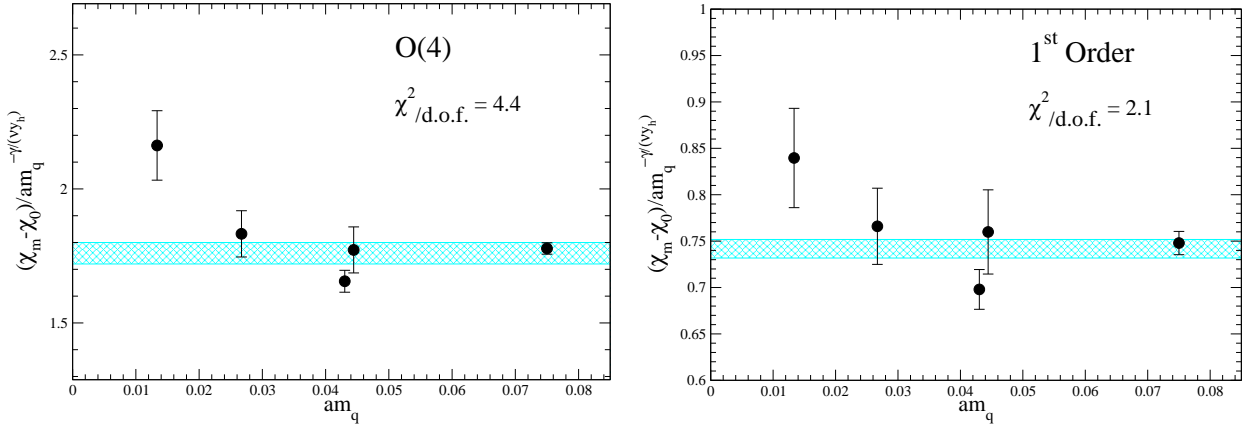

\includegraphics*[width=0.49\textwidth]{Chi_max_O4_2.eps}\hfill
\includegraphics*[width=0.49\textwidth]{Chi_max_1st_2.eps}
{%
\renewcommand{\thefigure}{\ref{SHCHI}$'$}%
\caption{\label{SHCHIp}Scaling of the maxima of $\chi_m-\chi_0$ at small $am_q$ for $O(4)$ (left) and first order (right).The value of $\chi_0$ is obtained by a best fit procedure. The $\chi^2/d.o.f$ of the best fit is also shown.}%
}
\end{figure*}

\begin{figure*}[tb!]
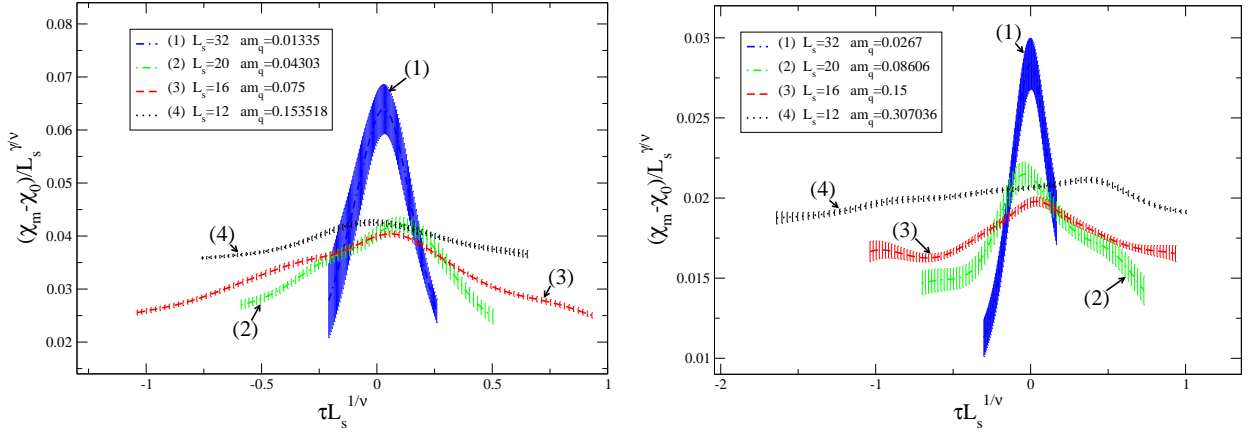

\includegraphics*[width=0.49\textwidth]{Chi_Run1_2.eps}\hfill
\includegraphics*[width=0.49\textwidth]{Chi_Run2_2.eps}
{%
\renewcommand{\thefigure}{\ref{R12SCA}$'$}%
\caption{\label{R12SCAp}Scaling of subtracted chiral condensate $\chi_m-\chi_0$ for Run1 (left) and Run2 (right). The background value $\chi_0$ is obtained from a best fit of the peak value.}%
}
\end{figure*}

\begin{figure*}[tb!]
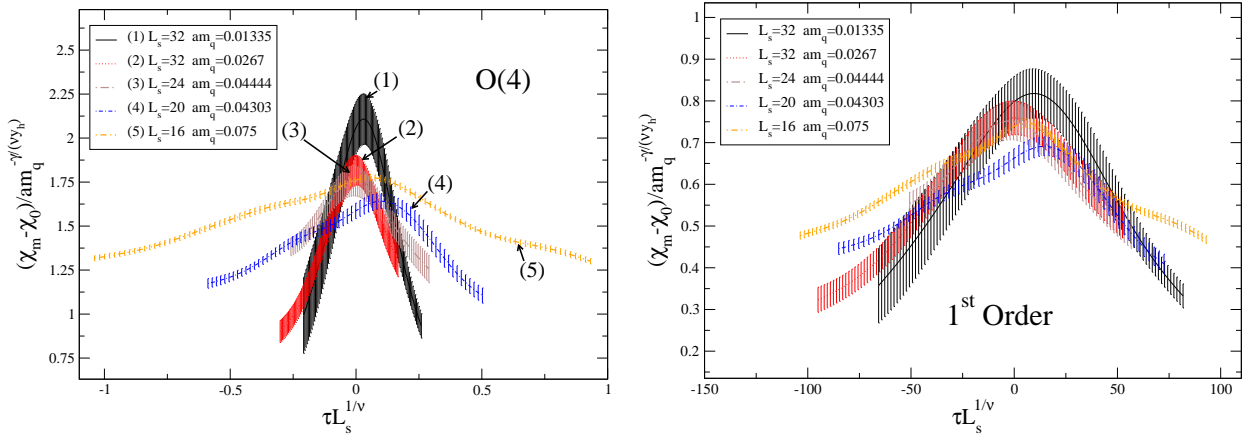

\includegraphics*[width=0.49\textwidth]{Chi_O4-Ls_2.eps}\hfill
\includegraphics*[width=0.49\textwidth]{Chi_1st_2.eps}
{%
\renewcommand{\thefigure}{\ref{SHCHISCA}$'$}%
\caption{\label{SHCHISCAp}Scaling of subtracted chiral condensate $\chi_m-\chi_0$ at small $am_q$ for $O(4)$ (left) and first order (right). The background value $\chi_0$ is obtained from a best fit of the peak value.}%
}
\end{figure*}

The best fit to determine the background is done by requesting scaling
for the peaks of $\chi_m$. the result is shown in Fig.s~\ref{R12p} and
\ref{SHCHIp}. For $O(4)$ ---and $O(2)$--- no scaling is obtained. 
A reasonable scaling results for first order. The corresponding
modified scaling at all $\beta$'s is shown in Fig.s~\ref{R12SCAp} and
\ref{SHCHISCAp}.

\newcount\mycount
\mycount=\arabic{figure}

\setcounter{figure}{\mycount}

We notice that, due to the dependence of the temperature $T$ on $am_q$
and not only on $\beta$ the definition of
$\langle\bar\psi\psi\rangle=\frac{\partial \ln Z}{\partial m} |_T$ and
$\chi_m = \frac{\partial^2 \ln Z}{\partial m^2} |_T$ must be revised,
resulting in a combination of several terms analogous to what happens
for the specific heat.

Analogously to what has been done for the specific heat, we have not
refined consequently our analysis. Our investigation is exploratory:
our $L_t$ is not big enough, our action is not improved. We will
extensively use the correct definitions in the planned improved
version of this investigation.

\subsection{Metastabilities}

\begin{figure}[thb]
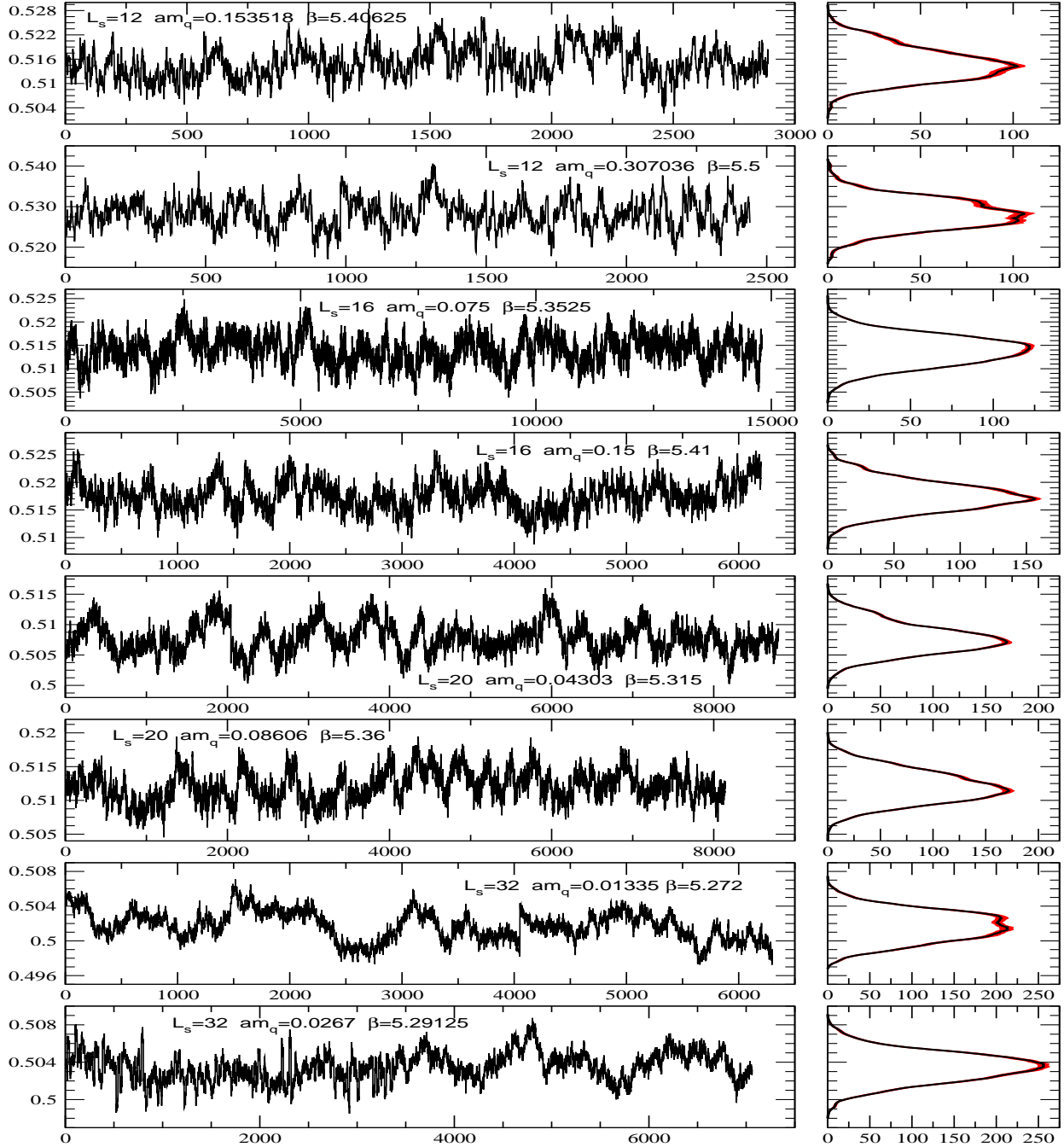

\includegraphics*[height=2.2cm,width=\columnwidth]{graph_12_04_5.40625_0.153518.eps} \\
\includegraphics*[height=2.2cm,width=\columnwidth]{graph_12_04_5.5_0.307036.eps} \\
\includegraphics*[height=2.2cm,width=\columnwidth]{graph_16_04_5.3525_0.075.eps} \\
\includegraphics*[height=2.2cm,width=\columnwidth]{graph_16_04_5.41_0.15.eps} \\
\includegraphics*[height=2.2cm,width=\columnwidth]{graph_20_04_5.315_0.04303.eps} \\
\includegraphics*[height=2.2cm,width=\columnwidth]{graph_20_04_5.36_0.08606.eps} \\
\includegraphics*[height=2.2cm,width=\columnwidth]{graph_32_04_5.272_0.01335.eps} \\
\includegraphics*[height=2.2cm,width=\columnwidth]{graph_32_04_5.29125_0.0267.eps} 
\caption{Probability distribution function of average plaquette. Chiral condensate shows a similar behavior.}\label{HIST}
\end{figure}

A first order chiral transition implies first order also at $m \neq 0$
and this should be visible in time histories. Metastable states should
be present and should be visible as double peak structures in the
histograms of distributions of the value of observables for large
enough volumes. Although the results for scaling presented in the
previous sections indicate that our volumes are not large enough, we
have analyzed the probability distribution function of a number of
observables, in particular of the spatial plaquettes. The results are
shown in Fig.~\ref{HIST}. As in previous works
\cite{fuku1,jlqcd} no convincing metastability appears.

To estimate the probability distribution function (PDF) from a given
finite data set a non-zero width of the integration region should be
chosen. We assume for the width a value given by one hundredth of the
difference between the largest and smallest entry in our
dataset. Errors are estimated using the bootstrap technique.

\section{Discussion and conclusions}\label{CONCLUSION}
The study of the nature of the chiral transition of $N_f=2$ QCD is of
fundamental importance: a second order phase transition would mean
crossover at $m\neq 0$ and also at finite chemical potential, implying
the presence of a tricritical point in the $T - \mu$ plane detectable
by heavy ion experiments; a first order phase transition would
drastically change this scenario. Moreover this is relevant to
understanding what confined and deconfined really means in Nature,
i.e. if they are two different phases of matter corresponding to
different realizations of some symmetry with an order parameter, or
are connected by a crossover.

Previous studies on the subject did not come to a definite conclusion,
mainly because of the huge computer power required.  We have
approached the problem by dedicating a big amount of computer power
and by proposing a novel strategy for the scaling study around the
chiral critical point. In particular we have developed a scaling study
which assumes the critical indexes of the expected second order
universality class ($O(4)$ or $O(2)$ according to the analysis of
Ref.~\cite{wilcz1}) to reduce to a one scale problem without any
further assumption. In this way we have been able to definitively rule
out the possibility of $O(4)$ or $O(2)$ critical indexes. 

We have introduced the mass dependence of the reduced temperature
$\tau$ (Eq.~\ref{taudef}), which was neglected in previous works.

We have then analyzed our data to test the scaling as done in previous
papers, assuming to be already in the thermodynamical limit
(Eq.~\ref{scalcal1} and \ref{scalord1}). The $am_q$ dependence of the
pseudocritical $\beta$ can not discriminate between $O(4)$, $O(2)$ and
first order. However the behavior of the peak of both the specific
heat and of the chiral susceptibility definitely excludes $O(4)$ and
$O(2)$, but are qualitatively consistent with first order
(Fig.~\ref{SHCHI} and \ref{SHCHIp}).

As for the shape of the critical peaks again $O(4)$ and $O(2)$ are
definitely ruled out. The dependence of the width on $L_s$ shows that
the thermodynamic limit is not reached. Instead a scaling at $L_s/\xi$
fixed agrees with first order, and again definitely excludes $O(4)$
and $O(2)$.

The magnetic equation of state is also nicely compatible with first
order.

No clear signal of metastability, except possibly some hint, is
observed.

In conclusion in $N_f=2$ QCD the chiral phase transition is not in the
universality class of $O(4)$. Data strongly indicate a first order
phase transition. Further study is needed to put this statement on a
firmer basis. First we are planning a similar analysis with improved
action and $L_t=6$. A consistency check will also be the study of the
$\eta'$ mass at the deconfining transition, in accordance with the
analysis of
\cite{wilcz1}, and we are also working at it.

\begin{acknowledgments}
We acknowledge useful discussion with L.~Del~Debbio, C.~DeTar,
E.~Laermann, B.~Lucini, G. Paffuti, P.~Petreczky.  We also acknowledge
F.~Karsch and A.~Ukawa for relevant remarks on a draft of the
manuscript.
\end{acknowledgments}

\appendix

\section{Monte Carlo parameters and raw data}\label{appA}
In this appendix we report the details of our numerical results. Time
histories of the observables are also available at request.

In the early stages of the computations some of the susceptibilities
were not measured. Blank entries in the following tables refer to
missing data. 

In some previous report the data for the plaquette susceptibilities
were defined with an extra factor of $1/4$. Below we have eliminated
it in agreement with definition Eq.~(\ref{LATSUSC3})~to~(\ref{LATSUSC7}).

The connected component $\chi_m^{conn}$ has been measured only at the
$\beta$ value nearest to the critical coupling (except for the run
with $L_s=24$, $am_q=0.04444$). 

\setlength{\LTcapwidth}{\textwidth}
\begin{longtable}{|c|c|c|c|c|c|c|c|}
\caption{Mean value of plaquettes, chiral condensate and energy density from MC simulation. The number of thermalized trajectories is also shown together with the extimated autocorrelation time (for $P_s$ and $\langle\bar\psi\psi\rangle$)}\\
\hline $\beta$ & \# Traj. & $P_\sigma$ & $P_\tau$ & $\langle\bar\psi\psi\rangle$ & $\langle\bar\psi D_0\psi\rangle$ & $\tau_{int}$ ($P_\sigma$) & $\tau_{int}$ ($\langle\bar\psi\psi\rangle$) \\
\hline
\endfirsthead
\caption{(continued)}\\
\hline $\beta$ & \# Traj. & $P_\sigma$ & $P_\tau$ & $\langle\bar\psi\psi\rangle$ & $\langle\bar\psi D_0\psi\rangle$ & $\tau_{int}$ ($P_\sigma$) & $\tau_{int}$ ($\langle\bar\psi\psi\rangle$) \\
\hline
\endhead
\hline\multicolumn{8}{|c|}{$L_t$=4 \hfil $L_s$=12 \hfil $am_q$=0.153518} \\
\hline 5.3 & 1300 &0.47880(16) &0.47893(16) &1.07450(60) &--- &7.7(2.4) &1.47(20) \\
\hline 5.325 & 1300 &0.48578(17) &0.48601(17) &1.04638(59) &--- &7.3(2.2) &1.71(26) \\
\hline 5.35 & 1300 &0.49337(21) &0.49363(20) &1.01300(79) &--- &10(3.7) &3.90(87) \\
\hline 5.375 & 2930 &0.50082(12) &0.50130(13) &0.97909(55) &0.72133(24) &9.2(2.0) &3.71(53) \\
\hline 5.375 & 1250 &0.50147(15) &0.50177(15) &0.97683(56) &--- &6.0(1.7) &1.91(31) \\
\hline 5.3875 & 2860 &0.50637(16) &0.50689(16) &0.95072(81) &0.72657(24) &12(3.2) &6.4(1.2) \\
\hline 5.3875 & 1200 &0.50653(30) &0.50711(32) &0.9497(20) &--- &18(9.3) &15(6.9) \\
\hline 5.4 & 2920 &0.51129(16) &0.51196(18) &0.9233(11) &0.73140(25) &9.7(2.2) &8.3(1.8) \\
\hline 5.4 & 1200 &0.51041(27) &0.51105(34) &0.9275(20) &--- &13(5.9) &16(7.7) \\
\hline 5.40625 & 2890 &0.51494(30) &0.51576(36) &0.8990(26) &0.73594(34) &32(14) &33(14) \\
\hline 5.40625 & 600 &0.51464(37) &0.51582(34) &0.8986(31) &--- &14(8.9) &16(10) \\
\hline 5.4125 & 2890 &0.51762(50) &0.51869(57) &0.8796(43) &0.74062(55) &72(46) &77(51) \\
\hline 5.4125 & 1200 &0.51955(36) &0.52100(49) &0.8635(37) &--- &22(12) &36(25) \\
\hline 5.41875 & 2900 &0.52200(36) &0.52357(40) &0.8470(28) &0.74866(41) &46(23) &42(20) \\
\hline 5.41875 & 840 &0.52247(49) &0.52409(53) &0.8468(37) &--- &20(13) &28(17) \\
\hline 5.425 & 2970 &0.52508(18) &0.52692(23) &0.8255(17) &0.75432(29) &13(3.5) &19(6.3) \\
\hline 5.425 & 1200 &0.52576(35) &0.52774(37) &0.8204(29) &--- &19(9.8) &20(11) \\
\hline 5.43125 & 2910 &0.52802(19) &0.52990(21) &0.8063(15) &0.75839(24) &17(5.4) &16(5.2) \\
\hline 5.43125 & 540 &0.52952(58) &0.53151(70) &0.7917(61) &--- &40(44) &66(90) \\
\hline 5.4375 & 1300 &0.53152(17) &0.53381(16) &0.7782(11) &--- &8.2(2.6) &8.6(2.8) \\
\hline 5.45 & 1300 &0.53475(20) &0.53730(22) &0.7598(16) &--- &10(3.9) &12(5.1) \\
\hline 5.475 & 1300 &0.54153(11) &0.54411(13) &0.72134(77) &--- &4.4(1.0) &5.7(1.5) \\
\hline 5.5 & 1300 &0.54614(10) &0.54917(11) &0.69862(54) &--- &4.7(1.1) &3.42(70) \\
\hline 5.525 & 1350 &0.55052(10) &0.55340(11) &0.67937(54) &--- &4.19(95) &3.41(70) \\
\hline 5.55 & 1350 &0.554344(89) &0.557310(76) &0.66412(41) &--- &3.13(62) &2.22(37) \\
\hline 5.575 & 1350 &0.558004(86) &0.560962(80) &0.65103(40) &--- &3.03(60) &2.14(35) \\
\hline 5.6 & 1350 &0.561446(72) &0.564575(86) &0.63826(36) &--- &2.32(41) &2.20(36) \\
\hline
\hline\multicolumn{8}{|c|}{$L_t$=4 \hfil $L_s$=16 \hfil $am_q$=0.075000} \\
\hline 5.3 & 670 &0.48822(16) &0.48859(20) &0.8911(34) &0.7461(19) &7.1(2.9) &11(6.3) \\
\hline 5.33 & 4840 &0.499282(83) &0.499991(91) &0.8124(14) &0.75303(77) &12(2.5) &24(7.2) \\
\hline 5.3325 & 3600 &0.50022(12) &0.50106(13) &0.8051(18) &0.75463(95) &18(5.2) &14(4.4) \\
\hline 5.335 & 4000 &0.50164(13) &0.50252(14) &0.7903(18) &0.75743(89) &19(5.3) &17(5.5) \\
\hline 5.3375 & 4000 &0.50321(12) &0.50418(14) &0.7779(20) &0.75825(81) &20(5.7) &35(13) \\
\hline 5.34 & 4000 &0.50425(22) &0.50530(24) &0.7667(23) &0.76050(84) &47(20) &48(21) \\
\hline 5.345 & 4000 &0.50778(20) &0.50905(23) &0.7337(24) &0.76576(81) &40(16) &47(21) \\
\hline 5.3475 & 4000 &0.51042(22) &0.51203(26) &0.7059(24) &0.77230(90) &40(16) &48(21) \\
\hline 5.35 & 7130 &0.51126(18) &0.51291(21) &0.6990(26) &0.77290(70) &40(12) &55(19) \\
\hline 5.3525 & 14800 &0.51397(16) &0.51585(19) &0.6664(22) &0.77862(49) &69(19) &78(23) \\
\hline 5.355 & 4000 &0.51614(35) &0.51826(41) &0.6446(42) &0.78217(90) &85(50) &91(55) \\
\hline 5.3575 & 3800 &0.51776(19) &0.52004(23) &0.6246(23) &0.78642(86) &36(14) &43(18) \\
\hline 5.36 & 13200 &0.51941(13) &0.52184(14) &0.6061(17) &0.78904(46) &52(13) &55(14) \\
\hline 5.365 & 4000 &0.52181(12) &0.52439(15) &0.5812(18) &0.79246(79) &21(6.5) &30(10) \\
\hline 5.37 & 4000 &0.52429(10) &0.52714(12) &0.5557(18) &0.79849(75) &17(4.5) &24(8.7) \\
\hline 5.38 & 3100 &0.527678(78) &0.530747(84) &0.5235(15) &0.80205(84) &9.9(2.2) &13(4.3) \\
\hline 5.4 & 1200 &0.532769(91) &0.53584(10) &0.4894(24) &0.8055(13) &5.1(1.3) &10(4.8) \\
\hline 5.45 & 1300 &0.542526(64) &0.545907(62) &0.4296(15) &0.8120(11) &3.14(64) &8.7(3.8) \\
\hline 5.5 & 1300 &0.550574(49) &0.553907(47) &0.3947(10) &0.81546(98) &2.24(40) &8.6(3.7) \\
\hline 5.6 & 775 &0.564279(58) &0.567819(47) &0.35047(99) &0.8180(15) &2.36(55) &31(28) \\
\hline
\hline\multicolumn{8}{|c|}{$L_t$=4 \hfil $L_s$=20 \hfil $am_q$=0.043030} \\
\hline 5.28 & 250 &0.48665(16) &0.48735(25) &0.8082(16) &--- &4.7(2.6) &5.4(2.7) \\
\hline 5.285 & 300 &0.48918(16) &0.48984(14) &0.7916(13) &--- &8.6(5.9) &3.0(1.2) \\
\hline 5.29 & 250 &0.49154(24) &0.49189(25) &0.7695(23) &--- &10(8.9) &7.9(5.2) \\
\hline 5.295 & 250 &0.49334(46) &0.49418(54) &0.7510(48) &--- &21(25) &17(17) \\
\hline 5.3 & 250 &0.49526(28) &0.49616(27) &0.7276(56) &--- &13(12) &22(23) \\
\hline 5.305 & 8640 &0.498846(93) &0.49999(10) &0.6983(11) &0.77044(10) &36(9.4) &31(7.7) \\
\hline 5.31 & 5366 &0.50211(15) &0.50351(17) &0.6598(23) &0.77587(15) &44(16) &48(18) \\
\hline 5.3125 & 2758 &0.50476(16) &0.50649(18) &0.6238(25) &0.78178(20) &27(11) &33(14) \\
\hline 5.315 & 8798 &0.50775(18) &0.50975(21) &0.5830(30) &0.78810(29) &69(24) &76(28) \\
\hline 5.3175 & 550 &0.50932(43) &0.51142(41) &0.5661(45) &0.78906(47) &42(46) &27(26) \\
\hline 5.32 & 6178 &0.51268(16) &0.51516(18) &0.5144(26) &0.79808(20) &43(14) &45(15) \\
\hline 5.325 & 200 &0.51513(27) &0.51791(36) &0.4769(33) &--- &18(22) &15(12) \\
\hline 5.33 & 300 &0.51894(17) &0.52218(20) &0.4298(30) &--- &7.8(5.1) &10(8.3) \\
\hline 5.34 & 350 &0.52239(26) &0.52557(33) &0.3933(43) &--- &19(17) &22(22) \\
\hline 5.35 & 400 &0.524922(98) &0.52834(10) &0.3677(15) &--- &4.3(1.8) &7.3(4.3) \\
\hline
\hline\multicolumn{8}{|c|}{$L_t$=4 \hfil $L_s$=32 \hfil $am_q$=0.013350} \\
\hline 5.24 & 350 &0.479565(70) &0.480086(81) &0.7687(18) &0.7631(10) &5.0(2.4) &3.4(1.6) \\
\hline 5.26 & 293 &0.48804(10) &0.48901(10) &0.6841(22) &0.7719(11) &13(11) &3.4(1.7) \\
\hline 5.27 & 1434 &0.49642(25) &0.49812(28) &0.5523(40) &0.78850(48) &78(74) &68(62) \\
\hline 5.2715 & 2323 &0.50038(47) &0.50252(52) &0.4783(98) &0.79708(59) &190(220) &190(230) \\
\hline 5.272 & 6300 &0.50173(22) &0.50405(25) &0.4539(47) &0.79952(33) &144(87) &147(90) \\
\hline 5.2725 & 3790 &0.50301(25) &0.50548(29) &0.4287(55) &0.80268(34) &106(71) &104(70) \\
\hline 5.2728 & 3175 &0.50422(18) &0.50685(21) &0.4019(42) &0.80561(27) &90(60) &86(57) \\
\hline 5.2731 & 2605 &0.50572(17) &0.50854(20) &0.3786(56) &0.80852(31) &77(53) &103(76) \\
\hline 5.27375 & 1060 &0.50511(34) &0.50792(41) &0.3859(65) &0.80829(42) &130(190) &100(130) \\
\hline 5.275 & 494 &0.50761(24) &0.51059(27) &0.3389(68) &0.81228(65) &40(47) &62(82) \\
\hline 5.28 & 335 &0.511175(49) &0.514484(69) &0.2679(12) &0.81864(63) &3.7(1.5) &8.1(5.0) \\
\hline 5.285 & 277 &0.51334(12) &0.51680(15) &0.2325(21) &0.82238(67) &14(13) &14(13) \\
\hline 5.29 & 290 &0.51485(13) &0.51841(15) &0.2097(19) &0.82267(54) &17(17) &13(12) \\
\hline 5.32 & 380 &0.522467(59) &0.526117(77) &0.14394(68) &0.82835(42) &4.2(1.8) &9.9(6.6) \\
\hline 5.4 & 295 &0.537482(40) &0.541084(38) &0.09389(20) &0.82998(44) &2.5(1.0) &3.5(1.9) \\
\hline
\hline\multicolumn{8}{|c|}{$L_t$=4 \hfil $L_s$=12 \hfil $am_q$=0.307036} \\
\hline 5.3 & 850 &0.46898(13) &0.46898(13) &1.24637(48) &--- &3.8(1.0) &0.94(13) \\
\hline 5.325 & 350 &0.47466(23) &0.47442(23) &1.23297(71) &--- &3.0(1.2) &1.11(26) \\
\hline 5.35 & 350 &0.48129(26) &0.48142(23) &1.21696(78) &--- &4.4(2.0) &1.39(37) \\
\hline 5.375 & 350 &0.48852(26) &0.48857(31) &1.19882(72) &--- &5.2(2.5) &1.67(45) \\
\hline 5.3875 & 350 &0.49244(40) &0.49255(43) &1.18851(97) &--- &9.3(6.2) &3.4(1.3) \\
\hline 5.4 & 350 &0.49571(29) &0.49575(33) &1.18102(72) &--- &7.0(4.1) &1.45(37) \\
\hline 5.4125 & 350 &0.49845(43) &0.49852(42) &1.17334(68) &--- &14(12) &0.91(19) \\
\hline 5.425 & 850 &0.50247(23) &0.50249(24) &1.16307(55) &--- &8.6(3.5) &2.87(67) \\
\hline 5.4375 & 850 &0.50662(29) &0.50682(29) &1.15084(62) &--- &15(8.3) &3.49(91) \\
\hline 5.45 & 800 &0.51059(16) &0.51091(13) &1.14039(55) &--- &5.1(1.6) &2.09(45) \\
\hline 5.475 & 3540 &0.51922(18) &0.51976(21) &1.11176(58) &0.67391(17) &16(4.7) &8.2(1.6) \\
\hline 5.475 & 750 &0.51906(26) &0.51934(29) &1.11220(79) &--- &10(5.3) &5.1(1.6) \\
\hline 5.4875 & 3530 &0.52352(17) &0.52394(17) &1.09638(59) &0.67698(20) &10(2.4) &6.4(1.1) \\
\hline 5.4875 & 700 &0.52333(51) &0.52363(57) &1.0998(25) &--- &24(17) &26(19) \\
\hline 5.49375 & 3390 &0.52652(17) &0.52721(17) &1.08530(60) &0.68075(23) &13(3.5) &6.9(1.2) \\
\hline 5.5 & 2440 &0.52830(20) &0.52916(21) &1.07792(79) &0.68299(28) &10(2.7) &7.1(1.5) \\
\hline 5.5 & 600 &0.52814(26) &0.52902(28) &1.0775(12) &--- &6.6(2.8) &9.8(4.4) \\
\hline 5.50625 & 3370 &0.53098(16) &0.53192(19) &1.06766(83) &0.68686(30) &12(3.0) &10(2.2) \\
\hline 5.5125 & 3460 &0.53436(23) &0.53573(28) &1.0530(11) &0.69200(32) &24(8.5) &20(6.2) \\
\hline 5.5125 & 700 &0.53450(41) &0.53582(54) &1.0522(21) &--- &19(13) &24(17) \\
\hline 5.525 & 3550 &0.53894(18) &0.54074(19) &1.03237(56) &0.69851(16) &23(7.9) &8.4(1.6) \\
\hline 5.525 & 700 &0.54025(23) &0.54216(24) &1.02766(80) &--- &9.6(4.5) &5.6(1.8) \\
\hline 5.5375 & 3430 &0.542920(92) &0.544869(97) &1.01947(33) &0.70273(16) &7.4(1.4) &3.41(44) \\
\hline 5.5375 & 700 &0.54369(16) &0.54587(26) &1.01300(74) &--- &5.5(2.0) &5.8(1.9) \\
\hline 5.55 & 1010 &0.54635(17) &0.54804(22) &1.00677(57) &0.70557(29) &8.6(3.2) &3.20(74) \\
\hline 5.55 & 700 &0.54622(17) &0.54839(23) &1.00761(65) &--- &5.4(1.9) &4.2(1.2) \\
\hline 5.575 & 700 &0.55133(13) &0.55365(16) &0.99023(37) &--- &4.2(1.3) &1.69(30) \\
\hline 5.6 & 800 &0.555811(97) &0.55827(12) &0.97684(36) &--- &2.77(69) &1.73(31) \\
\hline
\hline\multicolumn{8}{|c|}{$L_t$=4 \hfil $L_s$=16 \hfil $am_q$=0.150000} \\
\hline 5.3 & 2050 &0.479369(57) &0.479526(69) &1.0705(14) &0.7156(10) &3.83(68) &12(4.2) \\
\hline 5.36 & 16640 &0.496833(32) &0.497128(34) &0.99318(58) &0.72174(33) &7.95(70) &12(1.7) \\
\hline 5.38 & 13550 &0.503516(45) &0.503963(47) &0.96041(68) &0.72496(37) &10(1.1) &15(2.4) \\
\hline 5.39 & 9000 &0.507338(76) &0.507820(82) &0.93911(92) &0.72802(47) &19(3.7) &22(4.8) \\
\hline 5.4 & 10000 &0.511931(89) &0.51270(10) &0.91145(87) &0.73379(45) &26(5.6) &30(7.0) \\
\hline 5.405 & 8550 &0.514155(98) &0.51508(10) &0.8973(10) &0.73680(53) &26(5.9) &28(7.3) \\
\hline 5.41 & 6200 &0.51756(18) &0.51875(20) &0.8711(16) &0.74304(61) &49(17) &64(27) \\
\hline 5.415 & 8800 &0.52114(12) &0.52265(14) &0.8458(11) &0.75012(48) &38(10) &42(13) \\
\hline 5.42 & 12300 &0.52368(11) &0.52543(13) &0.82691(95) &0.75409(39) &40(9.4) &49(12) \\
\hline 5.425 & 14800 &0.526255(80) &0.528209(93) &0.80812(78) &0.75826(36) &31(5.7) &37(7.7) \\
\hline 5.43 & 9000 &0.528538(79) &0.530666(88) &0.79081(77) &0.76213(39) &21(4.1) &25(5.4) \\
\hline 5.44 & 7300 &0.532536(72) &0.534865(86) &0.76518(94) &0.76726(50) &16(3.2) &26(7.2) \\
\hline 5.46 & 7350 &0.537987(39) &0.540635(40) &0.73047(74) &0.77395(45) &6.83(85) &15(3.4) \\
\hline 5.47 & 1200 &0.540548(88) &0.543324(97) &0.7165(16) &0.77842(93) &5.1(1.3) &8.9(4.8) \\
\hline 5.48 & 1250 &0.542541(75) &0.545410(79) &0.7047(16) &0.7778(10) &4.20(99) &11(5.4) \\
\hline 5.49 & 1200 &0.544427(63) &0.547232(66) &0.6980(15) &0.78019(95) &3.48(76) &12(5.5) \\
\hline 5.5 & 1300 &0.546411(75) &0.549302(69) &0.6877(16) &0.78193(90) &4.14(96) &9.0(4.7) \\
\hline 5.52 & 900 &0.549775(69) &0.552711(78) &0.6736(14) &0.78312(98) &3.08(75) &8.6(4.5) \\
\hline
\hline\multicolumn{8}{|c|}{$L_t$=4 \hfil $L_s$=20 \hfil $am_q$=0.086060} \\
\hline 5.3 & 400 &0.486532(88) &0.48694(10) &0.92367(65) &--- &4.4(1.8) &1.09(25) \\
\hline 5.32 & 350 &0.49252(12) &0.49281(12) &0.89110(68) &--- &5.4(2.8) &1.73(54) \\
\hline 5.34 & 750 &0.50022(20) &0.50092(20) &0.8372(17) &--- &17(11) &16(9.5) \\
\hline 5.345 & 750 &0.50303(23) &0.50394(23) &0.8135(21) &--- &26(20) &23(16) \\
\hline 5.35 & 250 &0.50594(15) &0.50699(17) &0.7839(25) &--- &4.7(2.7) &11(9.3) \\
\hline 5.355 & 700 &0.50821(20) &0.50933(26) &0.7689(30) &--- &17(11) &35(31) \\
\hline 5.36 & 8140 &0.51208(17) &0.51359(20) &0.7305(19) &0.76815(25) &76(29) &77(29) \\
\hline 5.365 & 8565 &0.51635(12) &0.51829(13) &0.6858(15) &0.77747(19) &47(14) &56(18) \\
\hline 5.37 & 5574 &0.51993(12) &0.52223(13) &0.6488(15) &0.78332(15) &43(15) &52(20) \\
\hline 5.375 & 800 &0.52249(15) &0.52486(17) &0.6248(34) &--- &14(7.7) &54(60) \\
\hline 5.38 & 350 &0.52401(20) &0.52640(30) &0.6104(30) &--- &11(8.4) &20(19) \\
\hline 5.385 & 300 &0.52644(15) &0.52918(22) &0.5879(30) &--- &7.4(4.7) &15(15) \\
\hline 5.4 & 400 &0.53095(11) &0.53410(10) &0.5493(11) &--- &6.1(3.1) &8.3(5.2) \\
\hline 5.42 & 900 &0.536151(62) &0.539203(52) &0.51330(51) &--- &4.0(1.1) &5.4(1.7) \\
\hline
\hline\multicolumn{8}{|c|}{$L_t$=4 \hfil $L_s$=32 \hfil $am_q$=0.026700} \\
\hline 5.26 & 149 &0.48400(21) &0.48464(20) &0.7814(24) &0.7605(10) &10(11) &7.2(5.9) \\
\hline 5.28 & 650 &0.492092(63) &0.493055(77) &0.7023(10) &0.76843(57) &8.1(3.7) &4.8(1.7) \\
\hline 5.285 & 378 &0.49588(13) &0.49716(14) &0.6539(14) &0.77561(65) &15(13) &9.8(6.2) \\
\hline 5.29 & 1200 &0.50050(12) &0.50216(15) &0.5884(18) &0.78489(37) &34(23) &32(21) \\
\hline 5.29125 & 3960 &0.50398(18) &0.50605(21) &0.5299(33) &0.79314(34) &98(61) &98(62) \\
\hline 5.2925 & 3450 &0.504096(99) &0.50616(11) &0.5301(16) &0.79319(24) &24(8.2) &24(8.4) \\
\hline 5.2925 & 3118 &0.50426(36) &0.50635(41) &0.5273(58) &0.79334(42) &230(250) &210(220) \\
\hline 5.29375 & 950 &0.50641(12) &0.50880(15) &0.4896(22) &0.79804(42) &15(7.9) &19(11) \\
\hline 5.295 & 345 &0.50941(18) &0.51219(22) &0.4374(41) &0.80598(68) &23(24) &45(68) \\
\hline 5.3 & 465 &0.512593(80) &0.515578(86) &0.3892(13) &0.81064(48) &8.8(5.0) &11(7.1) \\
\hline 5.305 & 235 &0.515550(85) &0.518729(87) &0.3448(11) &0.81585(54) &7.3(5.3) &6.8(4.9) \\
\hline 5.31 & 300 &0.517129(79) &0.52044(10) &0.3232(16) &0.81816(62) &5.9(3.4) &10(8.9) \\
\hline 5.32 & 265 &0.520284(41) &0.523703(41) &0.28789(81) &0.82037(46) &2.6(1.1) &8.6(5.5) \\
\hline 5.34 & 285 &0.525526(40) &0.529029(39) &0.24220(75) &0.82483(47) &2.36(90) &7.4(4.9) \\
\hline
\hline\multicolumn{8}{|c|}{$L_t$=4 \hfil $L_s$=16 \hfil $am_q$=0.013350} \\
\hline 5.267 & 9600 &0.49225(24) &0.49348(28) &0.6246(34) &0.77948(55) &112(49) &66(22) \\
\hline 5.269 & 6350 &0.49564(48) &0.49725(54) &0.5688(88) &0.78565(66) &170(110) &152(94) \\
\hline 5.271 & 6500 &0.49997(47) &0.50214(54) &0.4872(93) &0.79538(71) &110(57) &112(59) \\
\hline 5.272 & 12400 &0.50248(63) &0.50489(72) &0.434(13) &0.80179(71) &250(140) &240(140) \\
\hline 5.273 & 6800 &0.50188(69) &0.50421(79) &0.449(13) &0.79947(79) &190(130) &170(110) \\
\hline 5.274 & 9350 &0.50519(50) &0.50795(58) &0.388(12) &0.80678(64) &180(100) &200(120) \\
\hline 5.276 & 2700 &0.50718(46) &0.51010(53) &0.343(11) &0.81174(81) &67(42) &72(49) \\
\hline 5.278 & 4200 &0.50995(16) &0.51318(19) &0.2863(42) &0.81673(55) &27(8.8) &42(17) \\
\hline 5.28 & 4200 &0.51161(14) &0.51496(15) &0.2562(31) &0.81963(53) &30(10) &35(13) \\
\hline
\hline\multicolumn{8}{|c|}{$L_t$=4 \hfil $L_s$=24 \hfil $am_q$=0.044440} \\
\hline 5.3 & 1310 &0.49537(14) &0.49619(16) &0.7421(17) &0.76363(20) &22(11) &20(10) \\
\hline 5.31 & 2200 &0.50213(20) &0.50355(23) &0.6645(29) &0.77583(24) &41(22) &42(23) \\
\hline 5.312 & 1990 &0.50368(24) &0.50518(27) &0.6465(35) &0.77825(24) &56(38) &57(38) \\
\hline 5.314 & 2620 &0.50495(17) &0.50663(18) &0.6310(27) &0.78066(26) &38(18) &42(21) \\
\hline 5.316 & 2520 &0.50750(38) &0.50945(44) &0.5953(63) &0.78634(49) &150(150) &160(170) \\
\hline 5.318 & 2600 &0.51024(22) &0.51249(25) &0.5554(35) &0.79230(30) &59(36) &63(39) \\
\hline 5.32 & 2260 &0.51128(29) &0.51360(31) &0.5430(43) &0.79402(31) &89(71) &92(75) \\
\hline 5.325 & 1470 &0.51599(17) &0.51879(18) &0.4773(28) &0.80338(18) &28(15) &30(17) \\
\hline 5.33 & 1420 &0.51875(10) &0.52173(10) &0.4426(15) &0.80829(15) &17(7.7) &20(9.5) \\
\hline 5.34 & 1000 &0.522071(90) &0.52527(10) &0.4049(16) &0.81235(17) &13(6.2) &26(16) \\
\hline 5.35 & 900 &0.525419(69) &0.528794(74) &0.37128(93) &0.81652(17) &7.4(2.7) &11(4.9) \\
\hline
\end{longtable}

\setlength{\LTcapwidth}{\textwidth}
\begin{longtable}{|c|c|c|c|c|c|c|c|}
\caption{Raw data from MC simulation of susceptibilities entering the specific heat.}\\
\hline $\beta$ & \# Traj. & $\chi_{e,\sigma\sigma}$ & $\chi_{e,\sigma\tau}$ & $\chi_{e,\tau\tau}$ & $\chi_{e,f}$ & $\chi_{e,\sigma}$ & $\chi_{e,\tau}$ \\
\hline
\endfirsthead
\caption{(continued)}\\
\hline $\beta$ & \# Traj. & $\chi_{e,\sigma\sigma}$ & $\chi_{e,\sigma\tau}$ & $\chi_{e,\tau\tau}$ & $\chi_{e,f}$ & $\chi_{e,\sigma}$ & $\chi_{e,\tau}$ \\
\hline
\endhead
\hline\multicolumn{8}{|c|}{$L_t$=4 \hfil $L_s$=12 \hfil $am_q$=0.153518} \\
\hline 5.3 & 1300 &0.0584(40) &0.0407(40) &0.0619(44) &--- &--- &--- \\
\hline 5.325 & 1300 &0.0570(38) &0.0375(35) &0.0559(38) &--- &--- &--- \\
\hline 5.35 & 1300 &0.0638(45) &0.0416(38) &0.0579(36) &--- &--- &--- \\
\hline 5.375 & 2930 &0.0598(27) &0.0426(28) &0.0625(32) &1.186(30) &0.0149(57) &0.0432(60) \\
\hline 5.375 & 1250 &0.0530(36) &0.0321(31) &0.0510(34) &--- &--- &--- \\
\hline 5.3875 & 2860 &0.0746(43) &0.0588(44) &0.0817(46) &1.174(30) &0.0424(78) &0.0741(88) \\
\hline 5.3875 & 1200 &0.0711(86) &0.0577(90) &0.0821(96) &--- &--- &--- \\
\hline 5.4 & 2920 &0.0877(52) &0.0732(58) &0.0962(61) &1.316(34) &0.086(10) &0.121(12) \\
\hline 5.4 & 1200 &0.0785(78) &0.0626(95) &0.084(10) &--- &--- &--- \\
\hline 5.40625 & 2890 &0.1094(83) &0.100(10) &0.129(12) &1.508(41) &0.160(19) &0.209(25) \\
\hline 5.40625 & 600 &0.067(10) &0.050(10) &0.0705(97) &--- &--- &--- \\
\hline 5.4125 & 2890 &0.134(13) &0.124(14) &0.155(16) &1.640(52) &0.213(31) &0.269(34) \\
\hline 5.4125 & 1200 &0.0848(90) &0.069(10) &0.095(12) &--- &--- &--- \\
\hline 5.41875 & 2900 &0.1128(85) &0.100(10) &0.125(11) &1.477(42) &0.155(21) &0.207(24) \\
\hline 5.41875 & 840 &0.105(13) &0.090(13) &0.112(15) &--- &--- &--- \\
\hline 5.425 & 2970 &0.0926(55) &0.0812(63) &0.1089(73) &1.370(36) &0.126(13) &0.173(17) \\
\hline 5.425 & 1200 &0.0824(92) &0.0654(94) &0.085(10) &--- &--- &--- \\
\hline 5.43125 & 2910 &0.0811(49) &0.0662(49) &0.0899(56) &1.221(32) &0.0654(92) &0.078(10) \\
\hline 5.43125 & 540 &0.0614(75) &0.0483(96) &0.069(11) &--- &--- &--- \\
\hline 5.4375 & 1300 &0.0521(36) &0.0344(34) &0.0532(34) &--- &--- &--- \\
\hline 5.45 & 1300 &0.0579(44) &0.0420(48) &0.0622(61) &--- &--- &--- \\
\hline 5.475 & 1300 &0.0425(24) &0.0265(24) &0.0452(29) &--- &--- &--- \\
\hline 5.5 & 1300 &0.0391(21) &0.0227(18) &0.0393(22) &--- &--- &--- \\
\hline 5.525 & 1350 &0.0379(19) &0.0222(17) &0.0415(21) &--- &--- &--- \\
\hline 5.55 & 1350 &0.0338(16) &0.0172(13) &0.0327(15) &--- &--- &--- \\
\hline 5.575 & 1350 &0.0320(16) &0.0152(11) &0.0287(13) &--- &--- &--- \\
\hline 5.6 & 1350 &0.0300(14) &0.0159(11) &0.0328(14) &--- &--- &--- \\
\hline
\hline\multicolumn{8}{|c|}{$L_t$=4 \hfil $L_s$=16 \hfil $am_q$=0.075000} \\
\hline 5.3 & 670 &0.0627(64) &0.0427(68) &0.0618(68) &2.09(47) &-0.027(64) &-0.019(84) \\
\hline 5.33 & 4840 &0.0732(33) &0.0573(33) &0.0808(36) &2.03(20) &0.040(27) &0.070(26) \\
\hline 5.3325 & 3600 &0.0835(51) &0.0678(55) &0.0911(60) &2.36(26) &0.051(31) &0.069(32) \\
\hline 5.335 & 4000 &0.0939(53) &0.0784(57) &0.1042(62) &2.26(22) &0.122(42) &0.140(41) \\
\hline 5.3375 & 4000 &0.0790(52) &0.0660(58) &0.0942(65) &1.82(20) &0.073(25) &0.104(24) \\
\hline 5.34 & 4000 &0.114(10) &0.102(11) &0.130(11) &1.98(21) &0.166(43) &0.216(39) \\
\hline 5.345 & 4000 &0.1090(82) &0.0980(96) &0.126(10) &1.82(19) &0.105(33) &0.137(35) \\
\hline 5.3475 & 4000 &0.126(11) &0.120(13) &0.152(15) &2.26(22) &0.231(47) &0.273(47) \\
\hline 5.35 & 7130 &0.1415(88) &0.135(10) &0.169(11) &2.40(18) &0.283(40) &0.363(43) \\
\hline 5.3525 & 14800 &0.1549(90) &0.150(10) &0.186(11) &2.40(13) &0.257(28) &0.335(35) \\
\hline 5.355 & 4000 &0.152(14) &0.148(17) &0.182(19) &2.27(24) &0.263(57) &0.321(63) \\
\hline 5.3575 & 3800 &0.1086(78) &0.1002(88) &0.129(10) &1.94(21) &0.182(40) &0.218(36) \\
\hline 5.36 & 13200 &0.1161(63) &0.1079(71) &0.1372(78) &1.93(10) &0.195(24) &0.243(27) \\
\hline 5.365 & 4000 &0.0806(62) &0.0706(72) &0.0978(84) &1.74(16) &0.104(30) &0.156(32) \\
\hline 5.37 & 4000 &0.0750(40) &0.0615(45) &0.0847(52) &1.58(15) &0.083(26) &0.123(31) \\
\hline 5.38 & 3100 &0.0552(30) &0.0395(29) &0.0605(32) &1.45(22) &0.074(25) &0.110(29) \\
\hline 5.4 & 1200 &0.0483(32) &0.0323(30) &0.0512(35) &1.43(29) &-0.053(47) &0.038(32) \\
\hline 5.45 & 1300 &0.0395(21) &0.0213(17) &0.0374(18) &1.19(18) &-0.040(35) &0.036(32) \\
\hline 5.5 & 1300 &0.0321(15) &0.0142(11) &0.0293(14) &0.81(21) &0.003(21) &0.039(23) \\
\hline 5.6 & 775 &0.0264(18) &0.0125(14) &0.0285(14) &1.19(30) &-0.030(27) &-0.000(34) \\
\hline
\hline\multicolumn{8}{|c|}{$L_t$=4 \hfil $L_s$=20 \hfil $am_q$=0.043030} \\
\hline 5.28 & 250 &0.069(10) &0.057(16) &0.093(22) &--- &--- &--- \\
\hline 5.285 & 300 &0.0531(76) &0.0307(78) &0.0479(74) &--- &--- &--- \\
\hline 5.29 & 250 &0.071(11) &0.052(10) &0.073(11) &--- &--- &--- \\
\hline 5.295 & 250 &0.114(46) &0.100(39) &0.126(40) &--- &--- &--- \\
\hline 5.3 & 250 &0.073(13) &0.054(10) &0.075(10) &--- &--- &--- \\
\hline 5.305 & 8640 &0.1101(55) &0.0982(60) &0.1256(67) &3.311(50) &0.144(13) &0.197(15) \\
\hline 5.31 & 5366 &0.151(11) &0.146(12) &0.181(14) &3.383(66) &0.246(28) &0.316(30) \\
\hline 5.3125 & 2758 &0.129(14) &0.116(16) &0.144(18) &3.056(87) &0.195(36) &0.252(44) \\
\hline 5.315 & 8798 &0.195(17) &0.196(19) &0.239(22) &3.440(82) &0.390(44) &0.481(51) \\
\hline 5.3175 & 550 &0.110(19) &0.096(20) &0.120(21) &2.50(20) &-0.017(63) &-0.020(70) \\
\hline 5.32 & 6178 &0.177(13) &0.175(14) &0.213(16) &2.869(64) &0.316(34) &0.403(38) \\
\hline 5.325 & 200 &0.0579(93) &0.041(14) &0.065(25) &--- &--- &--- \\
\hline 5.33 & 300 &0.0603(85) &0.0460(92) &0.067(12) &--- &--- &--- \\
\hline 5.34 & 350 &0.0724(88) &0.060(11) &0.080(13) &--- &--- &--- \\
\hline 5.35 & 400 &0.0444(57) &0.0262(49) &0.0462(48) &--- &--- &--- \\
\hline
\hline\multicolumn{8}{|c|}{$L_t$=4 \hfil $L_s$=32 \hfil $am_q$=0.013350} \\
\hline 5.24 & 350 &0.0639(67) &0.0455(78) &0.0642(93) &10(1.4) &0.03(10) &0.133(81) \\
\hline 5.26 & 293 &0.0551(78) &0.0394(73) &0.0628(77) &10(1.7) &0.06(10) &0.157(89) \\
\hline 5.27 & 1434 &0.205(40) &0.195(43) &0.226(47) &8.16(75) &0.38(11) &0.43(12) \\
\hline 5.2715 & 2323 &0.46(11) &0.50(13) &0.59(14) &9.87(74) &0.70(27) &0.77(28) \\
\hline 5.272 & 6300 &0.365(50) &0.394(54) &0.469(62) &8.59(36) &0.88(13) &1.04(15) \\
\hline 5.2725 & 3790 &0.371(60) &0.398(68) &0.471(78) &8.17(43) &0.83(16) &0.97(19) \\
\hline 5.2728 & 3175 &0.228(29) &0.231(32) &0.274(35) &6.23(32) &0.423(83) &0.498(92) \\
\hline 5.2731 & 2605 &0.184(25) &0.184(31) &0.226(35) &6.50(37) &0.322(74) &0.435(87) \\
\hline 5.27375 & 1060 &0.164(41) &0.168(49) &0.212(57) &5.62(54) &0.34(16) &0.41(16) \\
\hline 5.275 & 494 &0.123(27) &0.114(28) &0.145(31) &6.64(81) &0.234(97) &0.38(11) \\
\hline 5.28 & 335 &0.0479(55) &0.0318(56) &0.0517(56) &3.95(68) &-0.002(53) &0.119(65) \\
\hline 5.285 & 277 &0.0597(93) &0.050(10) &0.080(12) &3.79(59) &0.067(85) &0.085(68) \\
\hline 5.29 & 290 &0.065(14) &0.055(14) &0.080(19) &2.22(34) &0.057(50) &0.093(77) \\
\hline 5.32 & 380 &0.0567(50) &0.0369(58) &0.0591(71) &1.80(29) &-0.006(37) &0.019(35) \\
\hline 5.4 & 295 &0.0389(38) &0.0199(28) &0.0351(29) &1.50(27) &-0.049(28) &0.015(29) \\
\hline
\hline\multicolumn{8}{|c|}{$L_t$=4 \hfil $L_s$=12 \hfil $am_q$=0.307036} \\
\hline 5.3 & 850 &0.0468(29) &0.0248(25) &0.0447(30) &--- &--- &--- \\
\hline 5.325 & 350 &0.0577(73) &0.0342(59) &0.0552(57) &--- &--- &--- \\
\hline 5.35 & 350 &0.0552(52) &0.0336(50) &0.0483(52) &--- &--- &--- \\
\hline 5.375 & 350 &0.0483(59) &0.0290(60) &0.0545(64) &--- &--- &--- \\
\hline 5.3875 & 350 &0.069(12) &0.049(11) &0.067(11) &--- &--- &--- \\
\hline 5.4 & 350 &0.0495(71) &0.0323(75) &0.058(10) &--- &--- &--- \\
\hline 5.4125 & 350 &0.0536(64) &0.0382(61) &0.0596(69) &--- &--- &--- \\
\hline 5.425 & 850 &0.0611(51) &0.0423(51) &0.0647(60) &--- &--- &--- \\
\hline 5.4375 & 850 &0.0626(68) &0.0419(68) &0.0606(73) &--- &--- &--- \\
\hline 5.45 & 800 &0.0422(29) &0.0259(27) &0.0453(29) &--- &--- &--- \\
\hline 5.475 & 3540 &0.0816(58) &0.0655(61) &0.0871(66) &0.791(19) &0.0633(90) &0.091(10) \\
\hline 5.475 & 750 &0.0573(45) &0.0371(52) &0.0558(61) &--- &--- &--- \\
\hline 5.4875 & 3530 &0.0960(59) &0.0817(58) &0.1031(62) &0.830(21) &0.092(10) &0.121(10) \\
\hline 5.4875 & 700 &0.084(10) &0.072(10) &0.094(12) &--- &--- &--- \\
\hline 5.49375 & 3390 &0.0889(57) &0.0747(56) &0.0957(56) &0.845(21) &0.103(10) &0.128(11) \\
\hline 5.5 & 2440 &0.0902(60) &0.0792(67) &0.1025(78) &0.883(26) &0.099(11) &0.137(13) \\
\hline 5.5 & 600 &0.0652(65) &0.0482(73) &0.0655(76) &--- &--- &--- \\
\hline 5.50625 & 3370 &0.0846(47) &0.0729(52) &0.0960(57) &0.895(24) &0.1103(93) &0.143(10) \\
\hline 5.5125 & 3460 &0.0934(71) &0.0833(82) &0.1092(87) &0.877(26) &0.121(14) &0.159(15) \\
\hline 5.5125 & 700 &0.0703(80) &0.0588(94) &0.0807(98) &--- &--- &--- \\
\hline 5.525 & 3550 &0.0693(46) &0.0546(48) &0.0729(50) &0.695(16) &0.0615(77) &0.0881(84) \\
\hline 5.525 & 700 &0.0463(62) &0.0338(66) &0.0529(71) &--- &--- &--- \\
\hline 5.5375 & 3430 &0.0469(20) &0.0322(18) &0.0510(21) &0.611(14) &0.0227(35) &0.0475(37) \\
\hline 5.5375 & 700 &0.0412(31) &0.0300(33) &0.0536(43) &--- &--- &--- \\
\hline 5.55 & 1010 &0.0451(33) &0.0324(37) &0.0568(51) &0.611(26) &0.0298(66) &0.0569(87) \\
\hline 5.55 & 700 &0.0449(36) &0.0317(39) &0.0513(45) &--- &--- &--- \\
\hline 5.575 & 700 &0.0342(27) &0.0208(25) &0.0378(27) &--- &--- &--- \\
\hline 5.6 & 800 &0.0324(19) &0.0193(21) &0.0409(29) &--- &--- &--- \\
\hline
\hline\multicolumn{8}{|c|}{$L_t$=4 \hfil $L_s$=16 \hfil $am_q$=0.150000} \\
\hline 5.3 & 2050 &0.0484(20) &0.0277(21) &0.0495(23) &1.47(22) &0.015(31) &0.049(23) \\
\hline 5.36 & 16640 &0.0610(12) &0.0421(11) &0.0620(12) &1.235(65) &0.019(10) &0.048(10) \\
\hline 5.38 & 13550 &0.0702(17) &0.0519(17) &0.0721(18) &1.260(73) &0.038(14) &0.065(13) \\
\hline 5.39 & 9000 &0.0766(29) &0.0606(30) &0.0831(33) &1.345(91) &0.074(17) &0.090(19) \\
\hline 5.4 & 10000 &0.0860(34) &0.0722(38) &0.0969(43) &1.410(88) &0.072(16) &0.120(17) \\
\hline 5.405 & 8550 &0.0887(37) &0.0738(38) &0.0964(41) &1.57(11) &0.098(19) &0.149(22) \\
\hline 5.41 & 6200 &0.1156(86) &0.1066(97) &0.135(10) &1.52(13) &0.174(28) &0.215(35) \\
\hline 5.415 & 8800 &0.1012(62) &0.0918(66) &0.1196(73) &1.332(98) &0.127(21) &0.187(23) \\
\hline 5.42 & 12300 &0.1043(51) &0.0958(60) &0.1250(69) &1.289(81) &0.152(16) &0.207(19) \\
\hline 5.425 & 14800 &0.0880(33) &0.0758(37) &0.1002(41) &1.314(72) &0.126(14) &0.171(16) \\
\hline 5.43 & 9000 &0.0759(30) &0.0622(31) &0.0841(35) &1.205(81) &0.089(15) &0.120(17) \\
\hline 5.44 & 7300 &0.0663(28) &0.0523(31) &0.0739(35) &1.21(10) &0.065(16) &0.097(18) \\
\hline 5.46 & 7350 &0.0470(13) &0.0313(12) &0.0498(13) &0.972(81) &0.033(13) &0.062(14) \\
\hline 5.47 & 1200 &0.0440(27) &0.0285(24) &0.0463(28) &0.67(13) &0.046(29) &0.039(30) \\
\hline 5.48 & 1250 &0.0409(23) &0.0266(22) &0.0449(27) &0.88(16) &0.001(34) &0.027(28) \\
\hline 5.49 & 1200 &0.0355(19) &0.0200(17) &0.0391(20) &0.73(12) &0.030(34) &0.026(33) \\
\hline 5.5 & 1300 &0.0426(23) &0.0261(20) &0.0441(22) &0.66(12) &-0.012(25) &0.016(24) \\
\hline 5.52 & 900 &0.0334(21) &0.0157(17) &0.0324(22) &0.56(13) &-0.015(21) &0.002(22) \\
\hline
\hline\multicolumn{8}{|c|}{$L_t$=4 \hfil $L_s$=20 \hfil $am_q$=0.086060} \\
\hline 5.3 & 400 &0.0383(39) &0.0230(34) &0.0477(53) &--- &--- &--- \\
\hline 5.32 & 350 &0.0552(61) &0.0299(51) &0.0483(50) &--- &--- &--- \\
\hline 5.34 & 750 &0.086(11) &0.067(12) &0.085(12) &--- &--- &--- \\
\hline 5.345 & 750 &0.079(10) &0.063(11) &0.085(13) &--- &--- &--- \\
\hline 5.35 & 250 &0.0638(88) &0.0444(84) &0.0632(90) &--- &--- &--- \\
\hline 5.355 & 700 &0.078(12) &0.069(16) &0.096(18) &--- &--- &--- \\
\hline 5.36 & 8140 &0.166(12) &0.160(14) &0.194(15) &2.418(59) &0.047(27) &0.052(31) \\
\hline 5.365 & 8565 &0.1316(82) &0.1235(94) &0.154(10) &1.943(50) &0.050(18) &0.056(21) \\
\hline 5.37 & 5574 &0.1054(88) &0.0959(88) &0.1234(98) &1.670(40) &-0.013(14) &-0.012(16) \\
\hline 5.375 & 800 &0.0667(83) &0.0506(89) &0.0710(91) &--- &--- &--- \\
\hline 5.38 & 350 &0.0696(98) &0.054(10) &0.081(14) &--- &--- &--- \\
\hline 5.385 & 300 &0.0490(73) &0.0396(86) &0.067(10) &--- &--- &--- \\
\hline 5.4 & 400 &0.0435(53) &0.0255(40) &0.0389(40) &--- &--- &--- \\
\hline 5.42 & 900 &0.0400(27) &0.0209(22) &0.0366(24) &--- &--- &--- \\
\hline
\hline\multicolumn{8}{|c|}{$L_t$=4 \hfil $L_s$=32 \hfil $am_q$=0.026700} \\
\hline 5.26 & 149 &0.099(27) &0.076(25) &0.092(23) &5.6(1.2) &-0.11(14) &-0.10(13) \\
\hline 5.28 & 650 &0.0643(59) &0.0472(75) &0.0706(87) &5.65(78) &0.039(55) &0.123(64) \\
\hline 5.285 & 378 &0.081(23) &0.066(23) &0.090(26) &4.71(62) &0.034(77) &0.106(79) \\
\hline 5.29 & 1200 &0.110(17) &0.100(18) &0.133(22) &4.35(37) &0.219(54) &0.286(51) \\
\hline 5.29125 & 3960 &0.253(40) &0.261(44) &0.312(50) &5.46(29) &0.58(12) &0.69(13) \\
\hline 5.2925 & 3450 &0.286(21) &0.297(25) &0.351(28) &5.48(30) &0.589(65) &0.680(79) \\
\hline 5.2925 & 3118 &0.303(92) &0.31(10) &0.37(12) &5.74(41) &0.65(26) &0.82(30) \\
\hline 5.29375 & 950 &0.176(34) &0.175(35) &0.215(39) &4.56(45) &0.35(11) &0.41(12) \\
\hline 5.295 & 345 &0.102(19) &0.082(22) &0.100(25) &4.13(81) &0.127(73) &0.195(82) \\
\hline 5.3 & 465 &0.0706(85) &0.0512(85) &0.0725(96) &3.44(36) &-0.031(53) &0.012(45) \\
\hline 5.305 & 235 &0.0482(69) &0.0372(81) &0.0625(95) &2.16(36) &-0.030(48) &0.019(58) \\
\hline 5.31 & 300 &0.0666(90) &0.0431(96) &0.0552(89) &3.02(51) &-0.024(50) &0.046(47) \\
\hline 5.32 & 265 &0.0360(37) &0.0204(29) &0.0363(34) &2.04(32) &0.015(47) &0.036(47) \\
\hline 5.34 & 285 &0.0364(33) &0.0163(27) &0.0355(30) &1.96(37) &-0.091(44) &0.029(38) \\
\hline
\hline\multicolumn{8}{|c|}{$L_t$=4 \hfil $L_s$=16 \hfil $am_q$=0.013350} \\
\hline 5.267 & 9600 &0.148(15) &0.141(17) &0.175(20) &9.75(31) &0.259(53) &0.327(64) \\
\hline 5.269 & 6350 &0.242(45) &0.246(49) &0.294(56) &9.06(37) &0.41(11) &0.52(13) \\
\hline 5.271 & 6500 &0.310(40) &0.323(43) &0.380(49) &7.81(35) &0.63(11) &0.72(13) \\
\hline 5.272 & 12400 &0.445(52) &0.478(60) &0.558(67) &8.40(29) &0.95(13) &1.13(14) \\
\hline 5.273 & 6800 &0.377(48) &0.401(50) &0.470(56) &7.74(34) &0.75(12) &0.89(13) \\
\hline 5.274 & 9350 &0.313(49) &0.328(59) &0.386(68) &7.16(31) &-0.111(90) &-0.13(10) \\
\hline 5.276 & 2700 &0.193(34) &0.195(39) &0.235(44) &5.47(33) &0.301(72) &0.404(81) \\
\hline 5.278 & 4200 &0.1073(73) &0.0958(87) &0.1236(97) &4.20(19) &0.114(26) &0.185(29) \\
\hline 5.28 & 4200 &0.0819(78) &0.0669(82) &0.0904(87) &3.88(20) &0.089(27) &0.136(28) \\
\hline
\hline\multicolumn{8}{|c|}{$L_t$=4 \hfil $L_s$=24 \hfil $am_q$=0.044440} \\
\hline 5.3 & 1310 &0.108(13) &0.095(16) &0.124(18) &3.13(12) &0.138(26) &0.166(32) \\
\hline 5.31 & 2200 &0.171(17) &0.164(19) &0.198(21) &3.46(11) &0.304(41) &0.374(46) \\
\hline 5.312 & 1990 &0.174(22) &0.171(26) &0.211(29) &3.24(10) &0.306(55) &0.372(61) \\
\hline 5.314 & 2620 &0.156(21) &0.150(23) &0.187(25) &3.35(11) &0.295(56) &0.379(63) \\
\hline 5.316 & 2520 &0.187(39) &0.187(41) &0.230(49) &3.37(18) &0.38(12) &0.48(11) \\
\hline 5.318 & 2600 &0.170(18) &0.165(21) &0.199(24) &3.10(10) &0.333(49) &0.411(55) \\
\hline 5.32 & 2260 &0.184(28) &0.183(32) &0.225(32) &2.97(11) &0.335(68) &0.409(65) \\
\hline 5.325 & 1470 &0.128(19) &0.117(20) &0.143(21) &2.346(92) &0.189(45) &0.248(51) \\
\hline 5.33 & 1420 &0.0760(75) &0.0595(81) &0.0824(82) &1.883(75) &0.043(14) &0.084(16) \\
\hline 5.34 & 1000 &0.0579(74) &0.0419(78) &0.0615(86) &1.621(71) &0.038(12) &0.070(16) \\
\hline 5.35 & 900 &0.0556(42) &0.0365(41) &0.0561(43) &1.456(68) &0.018(11) &0.060(12) \\
\hline
\end{longtable}

\setlength{\LTcapwidth}{\textwidth}
\begin{longtable}{|c|c|c|c|c|c|}
\caption{Raw data from MC simulation for chiral condensate susceptibility and termal susceptibility.}\\
\hline $\beta$ & \# Traj. & $\chi_{m}^{disc}$ & $\chi_{t,\sigma}$ & $\chi_{t,\tau}$ & $\chi_{t,f}$\\
\hline
\endfirsthead
\caption{(continued)}\\
\hline $\beta$ & \# Traj. & $\chi_{m}^{disc}$ & $\chi_{t,\sigma}$ & $\chi_{t,\tau}$ & $\chi_{t,f}$\\
\hline
\endhead
\hline\multicolumn{6}{|c|}{$L_t$=4 \hfil $L_s$=12 \hfil $am_q$=0.153518} \\
\hline 5.3 & 1300 &0.859(32) &-0.183(21) &-0.204(21) &--- \\
\hline 5.325 & 1300 &0.806(31) &-0.180(20) &-0.173(18) &--- \\
\hline 5.35 & 1300 &0.863(37) &-0.238(23) &-0.227(22) &--- \\
\hline 5.375 & 2930 &0.947(27) &-0.238(15) &-0.241(17) &-0.077(42) \\
\hline 5.375 & 1250 &0.726(27) &-0.167(16) &-0.157(15) &--- \\
\hline 5.3875 & 2860 &1.077(43) &-0.336(27) &-0.362(27) &-0.302(53) \\
\hline 5.3875 & 1200 &1.17(11) &-0.353(62) &-0.397(69) &--- \\
\hline 5.4 & 2920 &1.396(67) &-0.468(35) &-0.510(43) &-0.639(81) \\
\hline 5.4 & 1200 &1.117(91) &-0.368(54) &-0.402(64) &--- \\
\hline 5.40625 & 2890 &1.95(17) &-0.730(77) &-0.810(95) &-1.37(18) \\
\hline 5.40625 & 600 &1.23(13) &-0.389(74) &-0.379(71) &--- \\
\hline 5.4125 & 2890 &2.31(22) &-0.91(10) &-1.00(12) &-1.81(23) \\
\hline 5.4125 & 1200 &1.43(20) &-0.504(89) &-0.57(10) &--- \\
\hline 5.41875 & 2900 &1.87(13) &-0.730(69) &-0.790(81) &-1.35(16) \\
\hline 5.41875 & 840 &1.51(17) &-0.62(11) &-0.67(12) &--- \\
\hline 5.425 & 2970 &1.585(97) &-0.588(45) &-0.669(52) &-1.09(12) \\
\hline 5.425 & 1200 &1.34(14) &-0.497(73) &-0.535(75) &--- \\
\hline 5.43125 & 2910 &1.316(75) &-0.296(31) &-0.324(35) &-0.697(81) \\
\hline 5.43125 & 540 &0.99(17) &-0.350(83) &-0.39(10) &--- \\
\hline 5.4375 & 1300 &0.709(42) &-0.230(23) &-0.245(23) &--- \\
\hline 5.45 & 1300 &0.859(67) &-0.291(32) &-0.324(38) &--- \\
\hline 5.475 & 1300 &0.486(30) &-0.157(15) &-0.178(17) &--- \\
\hline 5.5 & 1300 &0.384(16) &-0.1226(99) &-0.141(11) &--- \\
\hline 5.525 & 1350 &0.368(15) &-0.1287(90) &-0.1423(96) &--- \\
\hline 5.55 & 1350 &0.297(12) &-0.0956(76) &-0.1037(70) &--- \\
\hline 5.575 & 1350 &0.283(12) &-0.0885(74) &-0.0909(69) &--- \\
\hline 5.6 & 1350 &0.2442(92) &-0.0791(55) &-0.0944(61) &--- \\
\hline
\hline\multicolumn{6}{|c|}{$L_t$=4 \hfil $L_s$=16 \hfil $am_q$=0.075000} \\
\hline 5.3 & 670 &1.61(34) &-0.25(13) &-0.30(17) &-0.82(61) \\
\hline 5.33 & 4840 &1.82(21) &-0.381(58) &-0.425(60) &-0.20(27) \\
\hline 5.3325 & 3600 &2.20(25) &-0.481(82) &-0.570(88) &-0.56(38) \\
\hline 5.335 & 4000 &2.31(22) &-0.623(84) &-0.680(84) &-1.03(38) \\
\hline 5.3375 & 4000 &2.79(32) &-0.619(85) &-0.596(76) &-1.20(34) \\
\hline 5.34 & 4000 &3.89(37) &-1.09(12) &-1.24(13) &-1.77(39) \\
\hline 5.345 & 4000 &4.05(37) &-1.11(12) &-1.14(12) &-1.81(43) \\
\hline 5.3475 & 4000 &4.27(49) &-1.13(15) &-1.23(15) &-2.74(54) \\
\hline 5.35 & 7130 &5.56(48) &-1.48(15) &-1.67(16) &-3.65(51) \\
\hline 5.3525 & 14800 &5.75(42) &-1.58(12) &-1.85(14) &-3.57(37) \\
\hline 5.355 & 4000 &5.24(58) &-1.53(19) &-1.67(20) &-3.25(68) \\
\hline 5.3575 & 3800 &3.74(39) &-0.99(12) &-1.07(12) &-2.26(43) \\
\hline 5.36 & 13200 &4.29(32) &-1.162(97) &-1.35(10) &-2.60(29) \\
\hline 5.365 & 4000 &2.39(29) &-0.69(11) &-0.77(12) &-1.33(27) \\
\hline 5.37 & 4000 &2.26(24) &-0.606(77) &-0.743(91) &-1.06(31) \\
\hline 5.38 & 3100 &1.16(12) &-0.318(43) &-0.357(52) &-0.55(21) \\
\hline 5.4 & 1200 &1.16(27) &-0.226(97) &-0.294(82) &0.00(35) \\
\hline 5.45 & 1300 &0.501(95) &-0.116(40) &-0.184(43) &-0.09(24) \\
\hline 5.5 & 1300 &0.233(73) &-0.065(31) &-0.098(40) &-0.02(10) \\
\hline 5.6 & 775 &0.120(31) &-0.049(33) &-0.060(25) &0.17(13) \\
\hline
\hline\multicolumn{6}{|c|}{$L_t$=4 \hfil $L_s$=20 \hfil $am_q$=0.043030} \\
\hline 5.28 & 250 &3.16(35) &-0.43(11) &-0.64(17) &--- \\
\hline 5.285 & 300 &2.92(28) &-0.381(93) &-0.324(86) &--- \\
\hline 5.29 & 250 &3.60(40) &-0.56(12) &-0.53(12) &--- \\
\hline 5.295 & 250 &5.10(91) &-1.08(46) &-1.16(43) &--- \\
\hline 5.3 & 250 &5.4(1.1) &-0.66(18) &-0.70(16) &--- \\
\hline 5.305 & 8640 &5.27(24) &-1.109(72) &-1.227(80) &-2.25(17) \\
\hline 5.31 & 5366 &7.57(61) &-1.75(16) &-1.96(19) &-3.95(41) \\
\hline 5.3125 & 2758 &6.63(77) &-1.46(21) &-1.59(23) &-3.46(56) \\
\hline 5.315 & 8798 &11(1.1) &-2.62(28) &-2.98(32) &-6.65(73) \\
\hline 5.3175 & 550 &4.70(52) &-1.07(20) &-1.16(22) &0.38(80) \\
\hline 5.32 & 6178 &10.26(89) &-2.37(21) &-2.67(24) &-5.55(55) \\
\hline 5.325 & 200 &3.24(76) &-0.50(19) &-0.67(31) &--- \\
\hline 5.33 & 300 &3.02(60) &-0.58(13) &-0.65(18) &--- \\
\hline 5.34 & 350 &3.21(46) &-0.75(12) &-0.81(14) &--- \\
\hline 5.35 & 400 &1.39(17) &-0.293(64) &-0.325(60) &--- \\
\hline
\hline\multicolumn{6}{|c|}{$L_t$=4 \hfil $L_s$=32 \hfil $am_q$=0.013350} \\
\hline 5.24 & 350 &9.3(1.3) &-0.53(18) &-0.40(20) &-1(2.2) \\
\hline 5.26 & 293 &9.3(1.4) &-0.38(20) &-0.48(21) &-3(2.9) \\
\hline 5.27 & 1434 &17(2.8) &-2.76(63) &-3.00(73) &-7(1.6) \\
\hline 5.2715 & 2323 &50(10) &-6(2.3) &-7(2.6) &-23(5.0) \\
\hline 5.272 & 6300 &42(5.6) &-7(1.0) &-8(1.1) &-19(2.9) \\
\hline 5.2725 & 3790 &44(7.0) &-7(1.3) &-8(1.5) &-18(3.4) \\
\hline 5.2728 & 3175 &27(3.6) &-4.45(67) &-4.92(71) &-9(1.8) \\
\hline 5.2731 & 2605 &39(10) &-3.60(62) &-3.98(71) &-14(4.6) \\
\hline 5.27375 & 1060 &20(4.6) &-3.15(90) &-3(1.1) &-8(2.3) \\
\hline 5.275 & 494 &15(4.0) &-2.10(81) &-2(1.0) &-6(1.9) \\
\hline 5.28 & 335 &3.64(63) &-0.35(10) &-0.39(10) &-0.90(95) \\
\hline 5.285 & 277 &5.65(93) &-0.77(21) &-1.02(28) &-2(1.4) \\
\hline 5.29 & 290 &4.33(95) &-0.74(25) &-0.88(28) &-1.27(93) \\
\hline 5.32 & 380 &1.15(13) &-0.284(48) &-0.332(52) &0.21(38) \\
\hline 5.4 & 295 &0.083(14) &-0.046(14) &-0.042(15) &0.00(10) \\
\hline
\hline\multicolumn{6}{|c|}{$L_t$=4 \hfil $L_s$=12 \hfil $am_q$=0.307036} \\
\hline 5.3 & 850 &0.367(17) &-0.094(10) &-0.0813(96) &--- \\
\hline 5.325 & 350 &0.355(24) &-0.119(19) &-0.103(16) &--- \\
\hline 5.35 & 350 &0.373(28) &-0.092(17) &-0.097(18) &--- \\
\hline 5.375 & 350 &0.359(25) &-0.110(22) &-0.132(22) &--- \\
\hline 5.3875 & 350 &0.372(31) &-0.154(41) &-0.150(41) &--- \\
\hline 5.4 & 350 &0.367(26) &-0.115(25) &-0.103(28) &--- \\
\hline 5.4125 & 350 &0.320(21) &-0.089(18) &-0.090(18) &--- \\
\hline 5.425 & 850 &0.366(18) &-0.142(16) &-0.150(19) &--- \\
\hline 5.4375 & 850 &0.363(19) &-0.147(22) &-0.144(24) &--- \\
\hline 5.45 & 800 &0.350(18) &-0.094(12) &-0.110(12) &--- \\
\hline 5.475 & 3540 &0.591(21) &-0.250(22) &-0.270(25) &-0.147(34) \\
\hline 5.475 & 750 &0.387(24) &-0.152(19) &-0.154(22) &--- \\
\hline 5.4875 & 3530 &0.674(23) &-0.322(26) &-0.347(26) &-0.327(41) \\
\hline 5.4875 & 700 &0.594(75) &-0.305(51) &-0.334(54) &--- \\
\hline 5.49375 & 3390 &0.654(24) &-0.302(26) &-0.328(25) &-0.362(40) \\
\hline 5.5 & 2440 &0.675(31) &-0.333(28) &-0.364(32) &-0.377(48) \\
\hline 5.5 & 600 &0.430(42) &-0.176(27) &-0.180(29) &--- \\
\hline 5.50625 & 3370 &0.682(30) &-0.329(22) &-0.354(26) &-0.487(50) \\
\hline 5.5125 & 3460 &0.727(42) &-0.366(33) &-0.409(37) &-0.552(74) \\
\hline 5.5125 & 700 &0.488(45) &-0.254(35) &-0.284(44) &--- \\
\hline 5.525 & 3550 &0.513(17) &-0.234(19) &-0.246(21) &-0.185(29) \\
\hline 5.525 & 700 &0.352(25) &-0.142(28) &-0.169(29) &--- \\
\hline 5.5375 & 3430 &0.399(10) &-0.1341(85) &-0.1474(86) &-0.001(17) \\
\hline 5.5375 & 700 &0.317(18) &-0.108(12) &-0.148(17) &--- \\
\hline 5.55 & 1010 &0.337(16) &-0.126(12) &-0.146(17) &-0.094(31) \\
\hline 5.55 & 700 &0.295(16) &-0.128(14) &-0.147(17) &--- \\
\hline 5.575 & 700 &0.221(10) &-0.0641(83) &-0.0866(83) &--- \\
\hline 5.6 & 800 &0.2125(99) &-0.0642(72) &-0.0863(89) &--- \\
\hline
\hline\multicolumn{6}{|c|}{$L_t$=4 \hfil $L_s$=16 \hfil $am_q$=0.150000} \\
\hline 5.3 & 2050 &0.74(13) &-0.184(53) &-0.096(37) &0.28(25) \\
\hline 5.36 & 16640 &0.929(49) &-0.224(18) &-0.209(18) &-0.014(79) \\
\hline 5.38 & 13550 &1.062(69) &-0.320(27) &-0.317(27) &-0.21(10) \\
\hline 5.39 & 9000 &1.296(95) &-0.412(40) &-0.442(40) &-0.25(15) \\
\hline 5.4 & 10000 &1.295(80) &-0.431(34) &-0.497(38) &-0.72(13) \\
\hline 5.405 & 8550 &1.48(10) &-0.523(43) &-0.599(49) &-1.09(17) \\
\hline 5.41 & 6200 &1.94(23) &-0.758(93) &-0.80(10) &-1.55(29) \\
\hline 5.415 & 8800 &1.87(14) &-0.675(59) &-0.777(62) &-1.25(19) \\
\hline 5.42 & 12300 &1.88(11) &-0.672(50) &-0.793(53) &-1.18(14) \\
\hline 5.425 & 14800 &1.520(84) &-0.537(38) &-0.620(42) &-1.13(11) \\
\hline 5.43 & 9000 &1.180(77) &-0.417(36) &-0.461(41) &-0.72(11) \\
\hline 5.44 & 7300 &1.05(10) &-0.328(42) &-0.396(44) &-0.43(13) \\
\hline 5.46 & 7350 &0.657(52) &-0.235(24) &-0.231(23) &-0.18(10) \\
\hline 5.47 & 1200 &0.55(12) &-0.211(60) &-0.229(65) &-0.15(19) \\
\hline 5.48 & 1250 &0.526(99) &-0.256(53) &-0.207(46) &0.02(23) \\
\hline 5.49 & 1200 &0.49(10) &-0.194(59) &-0.171(54) &-0.18(21) \\
\hline 5.5 & 1300 &0.54(10) &-0.180(48) &-0.195(48) &0.07(17) \\
\hline 5.52 & 900 &0.283(70) &-0.100(30) &-0.106(44) &-0.06(11) \\
\hline
\hline\multicolumn{6}{|c|}{$L_t$=4 \hfil $L_s$=20 \hfil $am_q$=0.086060} \\
\hline 5.3 & 400 &1.384(93) &-0.145(28) &-0.184(37) &--- \\
\hline 5.32 & 350 &1.136(94) &-0.196(42) &-0.177(39) &--- \\
\hline 5.34 & 750 &2.04(29) &-0.56(11) &-0.58(11) &--- \\
\hline 5.345 & 750 &2.22(27) &-0.58(11) &-0.60(11) &--- \\
\hline 5.35 & 250 &2.31(27) &-0.49(11) &-0.521(98) &--- \\
\hline 5.355 & 700 &2.58(46) &-0.63(15) &-0.73(18) &--- \\
\hline 5.36 & 8140 &5.13(39) &-1.59(14) &-1.76(16) &-0.44(30) \\
\hline 5.365 & 8565 &4.16(31) &-1.24(10) &-1.38(11) &-0.59(22) \\
\hline 5.37 & 5574 &3.37(27) &-0.970(97) &-1.09(10) &0.07(19) \\
\hline 5.375 & 800 &2.00(40) &-0.52(14) &-0.58(14) &--- \\
\hline 5.38 & 350 &1.77(32) &-0.46(13) &-0.58(14) &--- \\
\hline 5.385 & 300 &1.86(38) &-0.42(12) &-0.57(15) &--- \\
\hline 5.4 & 400 &0.810(98) &-0.243(39) &-0.227(36) &--- \\
\hline 5.42 & 900 &0.649(40) &-0.164(18) &-0.171(17) &--- \\
\hline
\hline\multicolumn{6}{|c|}{$L_t$=4 \hfil $L_s$=32 \hfil $am_q$=0.026700} \\
\hline 5.26 & 149 &7.7(1.8) &-0.71(27) &-1.10(30) &0.6(2.4) \\
\hline 5.28 & 650 &4.38(57) &-0.409(96) &-0.46(13) &-1.52(93) \\
\hline 5.285 & 378 &5.9(1.0) &-0.76(33) &-0.80(32) &-1(1.3) \\
\hline 5.29 & 1200 &6.98(84) &-1.21(24) &-1.44(27) &-3.71(79) \\
\hline 5.29125 & 3960 &18(3.0) &-4.01(70) &-4.43(77) &-10(2.1) \\
\hline 5.2925 & 3450 &22(1.6) &-4.45(35) &-5.05(45) &-10(1.1) \\
\hline 5.2925 & 3118 &23(6.9) &-4(1.6) &-5(1.8) &-11(4.3) \\
\hline 5.29375 & 950 &13(2.7) &-2.54(58) &-2.99(65) &-6(2.0) \\
\hline 5.295 & 345 &5.9(1.2) &-1.13(24) &-1.26(30) &-3(1.2) \\
\hline 5.3 & 465 &3.91(55) &-0.72(13) &-0.56(11) &-0.34(67) \\
\hline 5.305 & 235 &2.28(42) &-0.46(11) &-0.42(11) &-0.80(63) \\
\hline 5.31 & 300 &3.13(70) &-0.60(18) &-0.57(18) &-0.49(65) \\
\hline 5.32 & 265 &1.55(25) &-0.236(85) &-0.304(80) &-0.64(50) \\
\hline 5.34 & 285 &1.23(18) &-0.282(55) &-0.259(65) &0.16(36) \\
\hline
\hline\multicolumn{6}{|c|}{$L_t$=4 \hfil $L_s$=16 \hfil $am_q$=0.013350} \\
\hline 5.267 & 9600 &14(1.1) &-2.08(29) &-2.30(36) &-4.85(93) \\
\hline 5.269 & 6350 &23(4.0) &-3.96(89) &-4(1.0) &-8(2.0) \\
\hline 5.271 & 6500 &30(4.1) &-5.35(80) &-5.96(90) &-13(2.2) \\
\hline 5.272 & 12400 &48(5.1) &-8(1.0) &-9(1.1) &-20(2.5) \\
\hline 5.273 & 6800 &42(5.0) &-7.19(98) &-8(1.0) &-17(2.6) \\
\hline 5.274 & 9350 &38(6.2) &1.25(63) &1.44(72) &-16(3.3) \\
\hline 5.276 & 2700 &23(4.8) &-3.84(84) &-4.30(95) &-8(1.6) \\
\hline 5.278 & 4200 &11.00(97) &-1.64(18) &-1.90(20) &-3.57(60) \\
\hline 5.28 & 4200 &7.84(90) &-1.15(17) &-1.27(18) &-2.51(58) \\
\hline
\hline\multicolumn{6}{|c|}{$L_t$=4 \hfil $L_s$=24 \hfil $am_q$=0.044440} \\
\hline 5.3 & 1310 &4.70(48) &-1.02(16) &-1.10(19) &-2.03(32) \\
\hline 5.31 & 2200 &8.53(86) &-2.03(24) &-2.24(27) &-4.81(66) \\
\hline 5.312 & 1990 &8.5(1.1) &-2.05(31) &-2.32(36) &-4.65(75) \\
\hline 5.314 & 2620 &8.6(1.1) &-1.95(31) &-2.20(34) &-5.12(85) \\
\hline 5.316 & 2520 &10(2.8) &-2.53(70) &-2.89(83) &-6(2.0) \\
\hline 5.318 & 2600 &9.5(1.1) &-2.21(29) &-2.45(33) &-5.70(76) \\
\hline 5.32 & 2260 &9.8(1.4) &-2.36(37) &-2.68(42) &-5.48(91) \\
\hline 5.325 & 1470 &7.1(1.2) &-1.63(30) &-1.76(33) &-3.46(76) \\
\hline 5.33 & 1420 &3.37(37) &-0.74(10) &-0.82(11) &-1.09(22) \\
\hline 5.34 & 1000 &2.43(40) &-0.50(11) &-0.55(12) &-0.86(26) \\
\hline 5.35 & 900 &1.56(14) &-0.365(50) &-0.412(50) &-0.70(11) \\
\hline
\end{longtable}

\begin{table}[b!]
\caption{Values of $\chi_m^{conn}$ merasured at the $\beta$ nearest to the pseudocritical coupling. This values was taken as a constant through the critical region and added to $\chi_m^{disc}$ to obtain $\chi_m$.}
\begin{tabular}{|c|c|c|c|c|}
\hline
$L_s$ & $am_q$ & $\beta$ & \# Traj. & $\chi_m^{conn}$ \\
\hline
\hline 12 & 0.153518 & 5.4125 & 5600 & 1.01(5)\\
\hline 16 & 0.075 & 5.35 & 5000 & 1.59(10)\\
\hline 20 & 0.04303 & 5.315 & 2500 & 2.12(5)\\
\hline 32 & 0.01335 & 5.2725 & 120\footnotemark[1] & 0.2(5)\footnotemark[1] \\
\hline 12 & 0.307036 & 5.5 & 6200 & 0.55(3)\\
\hline 16 & 0.15 & 5.41 & 5000 & 1.00(5)\\
\hline 20 & 0.08606 & 5.36 & 2550 & 1.3(1)\\
\hline 32 & 0.0267 & 5.2925 & 80\footnotemark[1] & 0.4(2)\footnotemark[1]\\
\hline 24 & 0.04444 & 5.316 & 3150 & 2.02(9)\\
\hline
\end{tabular}
\footnotetext[1]{This quantity was measured only on a small fraction of configurations. Due to limited statistics the resulting value for $\chi_m^{conn}$ is compatible with zero. However for these cases ($L_s=32$ and $am_q=0.01335, 0.0267$) $\chi_m^{conn}$ is only a small fraction of $\chi_m$ and can be safely neglected within errors.}
\end{table}

\begin{table}[b!]
\caption{Pseudocritical couplings $\beta_c$ extimated from the reweightes curves for $\chi_{e,\sigma\sigma}$.}
\begin{tabular}{|c|c|c|}
\hline
$L_s$ & $am_q$ & $\beta_c$\\
\hline
\hline 12 & 0.153518 & 5.4112(18)\\
\hline 16 & 0.075 & 5.35175(82)\\
\hline 20 & 0.04303 & 5.3164(11) \\
\hline 32 & 0.01335 & 5.27180(20)\\
\hline 12 & 0.307036 & 5.502(10)\\
\hline 16 & 0.15 & 5.41153(61)\\
\hline 20 & 0.08606 & 5.36072(77)\\
\hline 32 & 0.0267 & 5.29250(15)\\
\hline 24 & 0.04444 & 5.3164(18)\\
\hline 16 & 0.01335 & 5.27168(27)\\
\hline
\end{tabular}
\end{table}

\section{Estimate of the background of $C_V$}\label{appB}

No significant dependence of the background $C_0$ from the volume of
the system is expected since it is an ultraviolet quantity, while
dependence on $am_q$ and $\beta$ is expected. In order to estimate
$C_0(\beta,am_q)$ we performed a linear fit of the tails in the
$\beta$ region far from the peak for each different value of
$am_q$. The procedure used is the following:
\begin{itemize}
\item first estimate the width $W$ of the peak. The width that we choose to use as a reference is the width at $75\%$ of the total height of the peak;
\item then for each peak eliminate those points which lie in the $\beta$ region around the maximum of the curve at distances smaller than $n\cdot W$, where $n$ is a constant;
\item fit the remaining points with a linear function and study the dependence of this fitted background on the parameter $n$.
\end{itemize}
For the whole procedure to make sense, we require that the background
should fit better as $n$ is increased. We require the value to be
stable at sufficiently large $n$. This proves qualitatively to be the
case. Since the number of points at large $n$ is not large, we choose
the best value of $n$ by minimizing the reduced $\chi^2$ of the fit of
the background.

\begin{figure*}[b!]
\includegraphics*[width=0.95\textwidth]{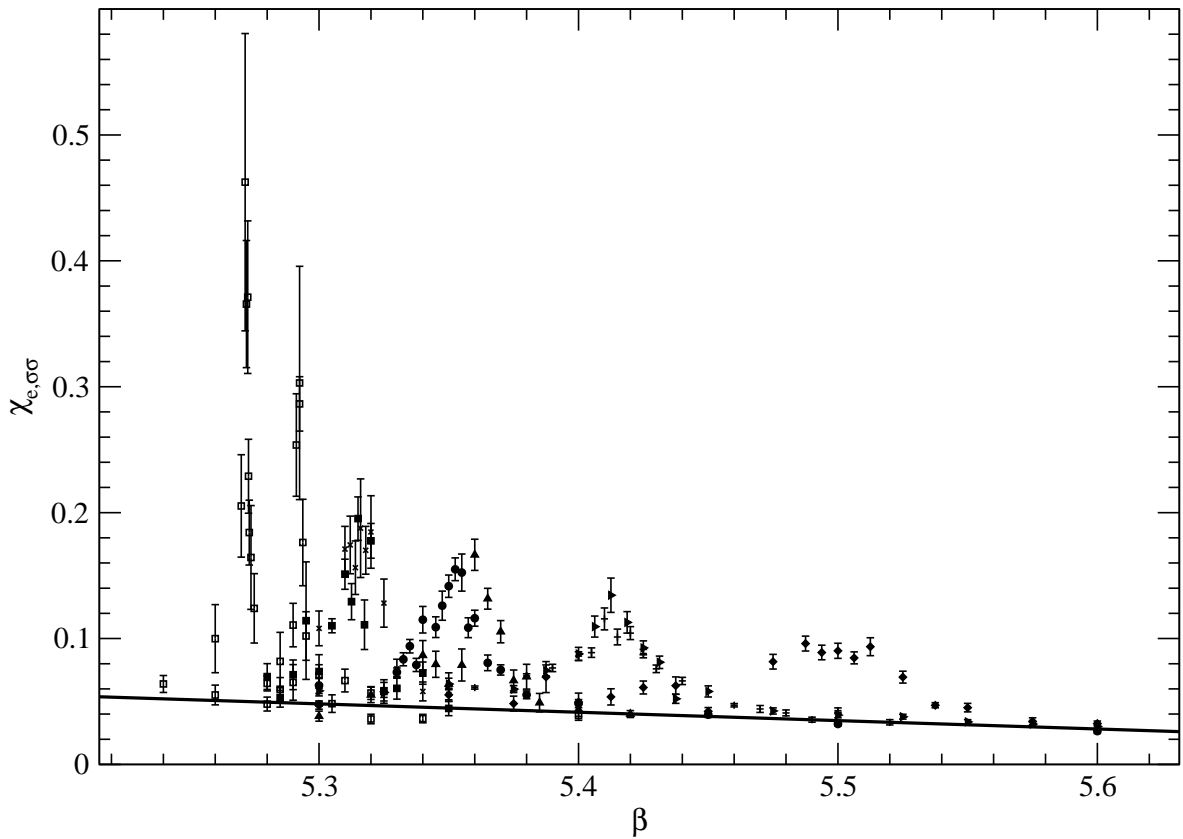}
\caption{Specific heat curves. The thick line shows the best linear fit for the background.}\label{CVFONDO}
\end{figure*}

No dependence of the background on the bare quark mass $am_q$ was
found so we can take $C_0(\beta,am_q)=C_0(\beta)$. We choose then to
do a global fit constraining all $C_V$ peaks at different masses to
share the same background. The result of this fit is shown in
Fig.\ref{CVFONDO}.
The best fit is obtained excluding a region of width
$12W$. Table~\ref{cvfitpar} shows the stability of the fit as the
parameter $n$ is in the range 6-15. 

It should be noticed that even the $\beta$ dependence is very weak and
that $C_0(\beta)$ is consistent with a constant in the $\beta$ range
examined within the statistical errors so that the whole procedure
described here is in practice equivalent to taking $C_0(\beta)$ a
constant.

\begin{table}[b!]
\caption{$C_0(\beta)$ fit parameters for different values of $n$. The fit function is $C_0(\beta) = q_1 + q_2 \beta$ }\label{cvfitpar}
\begin{tabular}{|c|c|c|c|c|c|}
\hline
$n$ & $q_1$ & $q_2$ & $\chi^2$/$d.o.f.$ & $d.o.f.$ \\ \hline \hline
3 & 0.608(60) & -0.103(11) & 8.70 & 84 \\ \hline
4 & 0.535(44) & -0.0908(84) & 5.17 & 77 \\ \hline
5 & 0.529(48) & -0.0887(88) & 5.19 & 66 \\ \hline
6 & 0.452(44) & -0.0751(80) & 3.56 & 58 \\ \hline
7 & 0.444(44) & -0.0740(80) & 3.62 & 53 \\ \hline
8 & 0.436(44) & -0.0728(80) & 3.61 & 47 \\ \hline
9 & 0.416(44) & -0.0693(77) & 3.19 & 42 \\ \hline
10 & 0.420(44) & -0.0695(75) & 3.11 & 37 \\ \hline
11 & 0.401(41) & -0.0661(77) & 2.89 & 34 \\ \hline
12 & 0.400(43) & -0.0663(83) & 2.81 & 29 \\ \hline
13 & 0.405(53) & -0.0668(97) & 2.99 & 26 \\ \hline
14 & 0.411(56) & -0.068(10) & 2.95 & 24 \\ \hline
15 & 0.404(56) & -0.067(10)  & 3.15 & 21 \\ \hline
\end{tabular}
\end{table}


\begin{thebibliography}{99}

\bibitem{Review} A recent review on QCD thermodynamics is contained in: 
E.~Laermann and O.~Philipsen, Ann.\ Rev.\ Nucl.\ Part.\ Sci.\  {\bf 53} (2003) 163 [arXiv:hep-ph/0303042].
\bibitem{wilcz1} R. D. Pisarski and F. Wilczek,  Phys. Rev. D {\bf 29}, 338 (1984); 
\bibitem{wilcz2} F. Wilczek, Int. J. Mod. Phys. A {\bf 7}, 3911 (1992); 
\bibitem{wilcz3} K. Rajagopal and F. Wilczek, Nucl. Phys. B {\bf 399}, 395 (1993)
\bibitem{Stephanov} M.~A.~Stephanov, K.~Rajagopal and E.~V.~Shuryak, Phys.\ Rev.\ Lett.\  {\bf 81} (1998) 4816 [arXiv:hep-ph/9806219].
\bibitem{fuku1} M. Fukugita, H. Mino, M. Okawa and A. Ukawa, Phys. Rev. Lett. {\bf 65}, 816 (1990).
\bibitem{fuku2} M. Fukugita, H. Mino, M. Okawa and A. Ukawa, Phys. Rev. D {\bf 42}, 2936 (1990).
\bibitem{colombia} F. R. Brown, F. P. Butler, H. Chen, N. H. Christ, Z. Dong, W. Schaffer, L. I. Unger and A. Vaccarino, Phys. Rev. Lett. {\bf 65}, 2491 (1990)
\bibitem{karsch1} F. Karsch, Phys. Rev. D {\bf 49}, 3791 (1994).
\bibitem{karsch2} F. Karsch and E. Laermann, Phys. Rev. D {\bf 50}, 6954 (1994).
\bibitem{jlqcd} S. Aoki et al. (JLQCD collaboration),  Phys. Rev. D {\bf 57}, 3910 (1998) [arXiv:hep-lat/9710048].
\bibitem{milc} C. Bernard, C. DeTar, S. Gottlieb, U. M. Heller, J. Hetrick, K. Rummukainen, R.L. Sugar and D. Toussaint,  Phys.Rev. D {\bf 61}, 054503 (2000)
\bibitem{cp-pacs} A. A. Khan et al. (CP-PACS collaboration), Phys. Rev. D {\bf 63}, 034502 (2001) [arXiv:hep-lat/0008011].
\bibitem{D'Elia:2004ua} M.~D'Elia, A.~Di Giacomo and C.~Pica, arXiv:hep-lat/0408008.
\bibitem{FISHER72} M.E.~Fisher and M.N.~Barber, Phys. Rev. Lett.{\bf 28} 1516 (1972)
\bibitem{BREZIN82} E.~Br\'ezin, J. Physique {\bf 43} 15 (1982)
\bibitem{HybridR} S.~A.~Gottlieb, W.~Liu, D.~Toussaint, R.~L.~Renken and R.~L.~Sugar, Phys.\ Rev.\ D {\bf 35} (1987) 2531.
\bibitem{massapi} T. Blum et al. (MILC collaboration), Phys. Rev. D {\bf 51} 5153 (1995).

\bibitem{newmanbarkema} M.E.J.~Newman and G.T.~Barkema, ``Monte Carlo Methods in Stastical Physics'', Oxford University Press (1999)
\bibitem{Karsch:1986ec}
F.~Karsch,
ILL-TH-86-9,
{\it Invited talk given at the Workshop on Lattice Gauge Theory - A Challenge in Large Scale Computing, Wuppertal, Germany, Nov 5-7, 1985}
\end{thebibliography}
\end{document}